\documentclass[11pt,twoside,a4paper,cmspaper,final,collab]{cms-tdr}
\def\svnVersion{f756983-D}\def\svnDate{2026/10/06}\def\cmsCernNoTag{CERN-EP-2025-192}\def\cmsCernDate{\today}\def\cmsMessage{Published in the Journal of High Energy Physics as \href{http://dx.doi.org/10.1007/JHEP10(2026)001}{\doi{10.1007/JHEP10(2026)001}.}}
\begin{document}\cmsNoteHeader{SMP-22-004}

\newcommand{\tunfold}	{\textsc{tunfold}\xspace}
\newcommand{\taup}      {\tau_{\perp}\xspace}
\newcommand{\mass}      {\rho_{\text{tot}}\xspace}
\newcommand{\massp}     {\rho^{{\mathrm T}}_{\text{tot}}\xspace}
\newcommand{\bT}        {B_{\text{tot}}\xspace}
\newcommand{\ytwo}      {Y_{23}\xspace}
\newcommand{\HTT}	{{H_{\mathrm{T,2}}}\xspace}
\newcommand{\nt}        {\widehat{n}}
\newcommand{\ptiv}      {{\vec{p}_{\mathrm{T},i}}} 
\newcommand{\pti}       {p_{\mathrm{T},i}}
\newcommand{\thrust}    {T_{\perp}}
\newlength\cmsTabSkip\setlength{\cmsTabSkip}{1ex}
\newcommand{\mgpy}{{\textsc{MG{5}}+\PYTHIAviii}\xspace}

\newlength\cmsFigWidth
\ifthenelse{\boolean{cms@external}}{\setlength\cmsFigWidth{0.49\textwidth}}{\setlength\cmsFigWidth{0.65\textwidth}} 

\cmsNoteHeader{SMP-22-004} 
\title{Measurement of event shape variables using charged particles inside jets in proton-proton collisions at \texorpdfstring{$\sqrt{s} = 13\TeV$}{sqrt(s) = 13 TeV}}

\date{\today}

\abstract{
Event shape variables, constructed from the four-momenta of the final-state objects in an event, are sensitive to the predictions of quantum chromodynamics in multijet production. A measurement of five event shape variables is presented, using proton-proton collision data collected at a centre-of-mass energy of 13\TeV with the CMS detector during 2016--2018, corresponding to an integrated luminosity of 138\fbinv. The variables are evaluated using the charged particles inside jets. After correcting for detector effects, their distributions are compared with the predictions from several models for multijet production. Overall, there is general agreement between several theoretical predictions and the data with scope for improvement of our understanding of QCD at different energy scales.
}
\hypersetup{%
pdfauthor={CMS Collaboration},%
pdftitle={Measurement of event shape variables using charged particles inside jets in proton-proton collisions at s = 13 TeV},%
pdfsubject={CMS},%
pdfkeywords={CMS, global features, jets}}

\maketitle 

\section{Introduction\label{sec:intro}}
The production of partons (quarks and gluons) in hadron collisions is important for the study of quantum chromodynamics (QCD), the theory of the strong interaction.
At higher energies, perturbative QCD calculations describe the production of partons. 
As a parton evolves into a set of stable hadrons, the energy involved in the process rapidly decreases and the use of phenomenological models becomes a necessity to describe
 parton shower (PS), fragmentation, and hadronization.
Hence, to validate the theoretical understanding of QCD, it is necessary to study variables that are sensitive to the entire process.
These models, implemented in different Monte Carlo (MC) simulation programs, known as  event generators, are improved and fine-tuned using experimental data.

Event shape variables (ESVs) are defined as ratios of hadron momenta or their combinations~\cite{Banfi:2010xy, Banfi:2004nk, Dasgupta:2003iq}.
They are sensitive to the topology of and energy flow in the final state of a high-energy collision event, and have been extensively studied in collider experiments.
The ESVs studied in this paper are infrared- and collinear-safe, and are robust against theoretical and experimental uncertainties~\cite{Banfi:2004nk, Dasgupta:2003iq}.
It has also been shown that the ESVs are sensitive to physics beyond the standard model~\cite{Chatterjee:2012qt, Datta:2011vg, Konar:2005bd}.

Previously, ESVs have been studied in electron-positron (\EE) collisions at PETRA (DESY)~\cite{PLUTO:1981inb, TASSO:1990cdg, Bethke:2009ehn} and LEP (CERN)~\cite{Abdallah:2003xz, Acciarri:1997dn, L3:2002oql, Achard:2004sv, OPAL:2011aa}, electron-proton ($\Pe\Pp$) collisions at HERA (DESY)~\cite{H1:1997hbl, H1:1999wfh, Aktas:2005tz, ZEUS:2002tyf, ZEUS:2006vwm}, and proton-antiproton ($\Pp\PAp$) collisions at the Tevatron (FNAL)~\cite{Aaltonen:2011et}. 
The results of these studies have been used to tune the parameters of the MC event generators, such as \PYTHIAviii and \HERWIG~\cite{Skands:2014pea, Abreu:1996na}, and also to estimate the value of the strong coupling \alpS~\cite{Dissertori:2007xa, Dissertori:2009ik, Jones:2003yv}.

Several studies of different ESVs have been performed at the CERN LHC using $\Pp\Pp$ collision data by the ATLAS~\cite{Aad:2012np, ATLAS:2017qir, ATLAS:2016hjr, ATLAS:2017qir, ATLAS:2020vup} and the CMS~\cite{Khachatryan:2011dx, Chatrchyan:2013tna, Khachatryan:2014ika, Sirunyan:2018adt} Collaborations, using jets as inputs.
Also, charged particles in the final states have been used to study ESVs by the ALICE~\cite{Abelev:2012sk, ALICE:2019dfi} and the ATLAS~\cite{Aad:2012fza} Collaborations.

These studies have shown varying degrees of agreement between data and the QCD models.
The evaluation of ESVs using jets has been shown to introduce theoretical issues related, \eg, to collinear and soft emissions of partons, or a dependence on the jet clustering; these issues may be reduced if individual particles are used to evaluate ESVs instead~\cite{Banfi:2010xy}.
In addition, the ESVs evaluated from jets become less sensitive to the hadronization process, an important aspect of QCD.
Using particles as inputs allows the testing and optimizing the hadronization models used in the simulations~\cite{Banfi:2010xy, Aad:2012fza}.

The high instantaneous luminosity during Run~2 of the LHC gave rise to, on average, 34 additional $\Pp\Pp$ interactions (pileup) per bunch crossing.
While pileup can be mitigated~\cite{CMS:2020ebo} at the level of individual jets, the same is not true for all the constituent particles.
However, a charged particle in an event is associated with a collision vertex, and the measurement of its energy and momentum is much less affected by pileup.
Hence, it is beneficial to use only the charged particles inside central jets, \ie, those within the acceptance of the tracking detector, to calculate the ESVs.

This paper describes a measurement of five ESVs: complement of transverse thrust ($\taup$), third jet resolution parameter ($\ytwo$), total jet broadening ($\bT$), total jet mass ($\mass$), and total transverse jet mass ($\massp$). 
The ESVs are evaluated with charged particles inside central jets in multijet events produced in $\Pp\Pp$ collisions at $\sqrt{s}=13\TeV$.
The distributions of the ESVs obtained from data are unfolded to correct for the detector effects.
The unfolded distributions are compared with the leading-order distributions predicted by the MC event generators \PYTHIAviii~\cite{Sjostrand:2014zea}, \HERWIG{}\,7~\cite{Bellm:2015jjp} and {\MGvATNLO}+\PYTHIAviii~\cite{Alwall:2014hca}. 
Tabulated results are provided in the HEPData record for this analysis~\cite{HEPData}.

\section{Event shape variables\label{sec:ESVs}}
This paper describes the study of five ESVs defined below.
Each of these variables has a value close to zero for events with back-to-back topologies, and a higher value for multijet, spherical events where the jets are distributed isotropically.
For hadron colliders, ESVs have been redefined using the input objects from the central region of the detector~\cite{Banfi:2004nk,Banfi:2010xy,Dasgupta:2003iq}. 
In this study, the four-momenta of the charged particles inside the selected jets (see Section~\ref{sec:evntsel}) are used to evaluate the ESVs.

\begin{itemize}
\item \textit{Complement of transverse thrust:} the complement of transverse thrust ($\taup$) is defined as
\begin{equation}\label{eqn:Complement}
\taup \equiv 1 - \thrust,
\end{equation}
where the event thrust observable in the transverse plane is defined as
\begin{equation}
\thrust \equiv \max_{\nt} \frac{\sum_i |\ptiv \cdot \nt |} {\sum_i \abs{\ptiv}} \label{eqn:Thrust},
\end{equation}
where $\ptiv$ is the transverse momentum of the $i^{\text{th}}$ object and $\nt$ is the unit vector, defined as the thrust direction, that maximizes the projection.
The limiting value of $\taup$ is zero for events with only two perfectly back-to-back jets and  ${(1 - 2/\pi)}$ for perfectly isotropic multijet events.

\item \textit{Third-jet resolution parameter:} the third-jet resolution parameter ($\ytwo$) defined as
\begin{equation} 
        \ytwo \equiv \frac{\max\left(p^{2}_{\mathrm{T},3},\left[\min(p_{\mathrm{T},i},p_{\mathrm{T},j})^2 (\Delta R_{i,j})^{2}/R^2\right]\right)}{p^{2}_{\mathrm{T},12}},
\label{eqn:y23}
\end{equation}

where $i$, $j$ run over all three final-state inputs and $(\Delta R_{i,j})^2 =(\eta_{i} - \eta_{j})^{2} + (\phi_{i} - \phi_{j})^{2}$. 

For the events containing more than three final-state particles, they are recombined with the distance parameter $R = 0.4$ until only three particles are left in the final state. 
The quantity $p_{\mathrm{T},3}$ denotes the transverse momentum of the third particle (pseudo-particle). 
The three particles (pseudo-particles) are further recombined into two as described above. 
The variable $p_{\mathrm{T},12}$ is then defined as the scalar \pt sum of these two remaining particles or pseudo-particles.

This variable approaches zero for ideal two-jet-like configurations, while nonzero values indicate the presence of additional hard radiation. 
Thus, $\ytwo$ provides sensitivity to hard parton emissions and serves as a probe of QCD parton shower modeling.

\item \textit{Jet broadening:} In each event, the thrust direction $\nt$ is used to divide the central region into two hemispheres.
All particles in the upper hemisphere (U) have ${\vec{p}}_\mathrm{T}\cdot\nt > 0$ and in the lower hemisphere (L), ${\vec{p}}_\mathrm{T}\cdot\nt < 0$. 
The pseudorapidities ($\eta$) and the azimuthal angles ($\phi$) of the axes for the upper and lower hemispheres are defined by
\begin{equation*}
	\eta_X \equiv \frac{\sum_{i\in{X}} p_{\mathrm{T},i}\,\eta_i} {\sum_{i\in{X}} p_{\mathrm{T},i}}, ~~
	\phi_X \equiv \frac{\sum_{i\in{X}} p_{\mathrm{T},i}\,\phi_i} {\sum_{i\in{X}} p_{\mathrm{T},i}},
\end{equation*}
where $X$ refers to the upper (U) or lower (L) hemisphere.
The jet broadening variable in each hemisphere is defined as
\begin{equation}
	B_{X} \equiv {\frac{1}{2\,P_{\mathrm{T}}} \sum_{i\in{X}}
p_{\mathrm{T},i} \sqrt{(\eta_i -\eta_X)^2 + (\phi_i - \phi_X)^2}},
\end{equation}
where ${P_\mathrm{T}}$ is the scalar $\pt$ sum of all the input objects.
The total jet broadening is then defined as $\bT \equiv  B_\mathrm{U} + B_\mathrm{L}$.

\item \textit{Total jet mass:} the normalized squared invariant mass of the jets in the upper and lower hemispheres of the event is defined by
\begin{equation}
\rho_X \equiv \frac{M^2_X}{P^2},
\end{equation}
where ${M_X}$ is the invariant mass of the constituents of the jets in the region $X$, and $P$ is the scalar sum of the momenta of all constituents in both hemispheres. The total jet mass is defined as the sum of the masses in the upper and lower regions,
\begin{equation}
	\mass \equiv \rho_\mathrm{U} + \rho_\mathrm{L}.     \label{eqn:JetMass}
\end{equation}

\item \textit{Total transverse jet mass:} the corresponding quantity of total jet mass in the transverse plane, the total transverse jet mass ($\massp$), is similarly calculated using the corresponding $\pt$.
\end{itemize}

According to the definitions, $\taup$ is more dependent on the initial hard scattering, the jet masses and jet broadening depend more on the subsequent 
nonperturbative part, \ie, the evolution of the partons into hadrons, and $\ytwo$ is sensitive to the 
clustering and substructure of the jets~\cite{Banfi:2010xy, Khachatryan:2014ika, Sirunyan:2018adt}.

\section{The CMS detector\label{sec:cms}}
The central feature of the CMS apparatus~\cite{CMS:2008xjf} is a superconducting solenoid of 6\unit{m} internal diameter, providing a magnetic field of 3.8\unit{T}.
The solenoid volume has a silicon pixel and strip tracker, a lead tungstate crystal electromagnetic calorimeter (ECAL), and a brass and scintillator hadron calorimeter (HCAL), each composed of a barrel and two endcap sections.
Forward calorimeters extend the pseudorapidity coverage provided by the barrel and endcap detectors.
Muons are measured in gas-ionization detectors embedded in the steel flux-return yoke outside the solenoid.

The tracker measures the charge and momentum of charged particles.
Until the end of 2016, it consisted of 1440 silicon pixel and 15\,148 silicon strip detector modules and covered the range $\abs{\eta} < 2.5$.
The upgraded tracker has 1856 silicon pixel and 15\,148 silicon strip detector modules with an extended coverage with $\abs{\eta} < 3.0$~\cite{Phase1Pixel}.
For nonisolated particles of $1 < \pt < 10\GeV$ and $\abs{\eta} < 3.0$, the track resolutions are typically 1.5\% in \pt and 20--75\mum in the transverse impact parameter~\cite{DP-2020-049}.

In the barrel section of the ECAL, energy resolution of about 1\% is achieved for unconverted or late-converting photons in the  tens of \GeVns energy range.
The energy resolution of the barrel photons that convert early is about 1.3\% up to $\abs{\eta} = 1$, rising to about 2.5\% at $\abs{\eta} = 1.4$. 
In the endcaps, the energy resolution is about 2.5\% for unconverted or late-converting photons, and between 3--4\% for the ones that convert early~\cite{CMS:2015myp}.
When combining information from the entire detector, the jet energy resolution amounts typically to 15--20\% at 30\GeV, 10\% at 100\GeV, and 5\% at 1\TeV~\cite{Khachatryan:2016kdb}.
In the region $\abs{\eta} < 1.74$, the HCAL cells have widths of 0.087 in both $\eta$ and $\phi$.
In the $\eta$-$\phi$ plane, and for $\abs{\eta} < 1.48$, the HCAL cells map on to $5{\times}5$ arrays of ECAL crystals to form calorimeter towers projecting radially outwards from close to the nominal interaction point.
For $\abs{\eta} > 1.74$, the coverage of the towers increases progressively to a maximum of 0.174 in $\Delta \eta$ and $\Delta \phi$.
Within each tower, the energy deposits in ECAL and HCAL cells are summed to define the calorimeter tower energies, which are subsequently used to provide the energies and directions of hadronic jets. 

Events of interest are selected using a two-tiered trigger system. The first level (L1), composed of custom hardware processors, uses information from the calorimeters and muon detectors to select events at a rate of around 100\unit{kHz} within a fixed latency of about 4\mus~\cite{CMS:2020cmk}.
The second level, known as the high-level trigger (HLT), consists of a farm of processors running a version of the full event reconstruction software optimized for fast processing, and reduces the event rate to a few \unit{kHz} before data storage~\cite{CMS:2016ngn,CMS:2024aqx}.
A more detailed description of the CMS detector can be found in Ref.~\cite{CMS:2008xjf,CMS:2023gfb}.

\section{Event reconstruction\label{sec:jetrec}}
A particle-flow algorithm~\cite{CMS:2017yfk} aims to reconstruct and identify each individual particle in an event, with an optimized combination of information from the various elements of the CMS detector.

Photons are identified as ECAL energy clusters not linked to the extrapolation of any charged-particle trajectory. An electron is identified as a primary charged particle track which is associated with energy clusters in the ECAL corresponding to its extrapolation to the ECAL and potential bremsstrahlung photons.
The energy of electrons is determined from a combination of the track momentum at the main interaction vertex, the corresponding ECAL cluster energy, and the energy sum of the bremsstrahlung photons.
Muons are identified as tracks in the central tracker consistent with either a track or several hits in the muon system, and associated with calorimeter deposits compatible with the muon hypothesis.
The momentum is obtained from the corresponding track curvature. 

A charged-particle track that is not identified as either an electron or muon gives rise to a charged hadron.
The energy is determined from a combination of the track momentum and the corresponding ECAL and HCAL energies.
Finally, neutral hadrons are identified as HCAL energy clusters not linked to any charged-hadron trajectory, or as a combined ECAL and HCAL energy excess with respect to the expected charged-hadron energy deposit.
The energy of neutral hadrons is obtained from the corresponding corrected ECAL and HCAL energies.

The $\Pp\Pp$ interaction vertices are reconstructed from hits in the inner tracking detectors. The primary vertex is taken to be the vertex corresponding to the 
scattering with highest scalar sum of \pt of the particles in the event, evaluated using tracking information alone, as described in Section 9.4.1 of Ref.~\cite{CMS-TDR-15-02}.

For each event, hadronic jets are clustered from these reconstructed particles using the infrared- and collinear-safe anti-\kt algorithm~\cite{Cacciari:2008gp, Cacciari:2011ma} with a distance parameter of 0.4.
Jet momentum is determined as the vectorial sum of all particle momenta in the jet, and is found from simulation to be, on average, within 5--10\% of the true momentum over the entire \pt spectrum and detector acceptance. 
Pileup interactions can contribute additional tracks and calorimetric energy depositions to the jet momentum. 
To mitigate this effect, tracks identified to be originating from pileup vertices are discarded and an offset correction is applied to correct for the remaining contributions~\cite{CMS:2020ebo}.

Jet energy corrections are derived from simulation to bring the measured response of jets to that of particle-level jets on average. In situ measurements of the momentum balance in dijet, $\text{photon}+\text{jet}$, $\PZ+\text{jet}$, and multijet events are used to account for any residual 
differences in the jet energy scale between data and simulation~\cite{Khachatryan:2016kdb}. 
The jet energy resolution amounts typically to 15--20\% at 30\GeV, 10\% at 100\GeV, and 5\% at 1\TeV~\cite{Khachatryan:2016kdb}. Additional selection criteria are applied to each jet to remove jets potentially dominated by instrumental effects or reconstruction failures~\cite{CMS:2020ebo}.

\section{Data and simulated events\label{sec:datasim}}
\subsection{Data sample}
This analysis uses $\Pp\Pp$ collision data at $\sqrt{s} = 13\TeV$ collected by the CMS experiment during Run~2 (2016--2018) of the LHC, corresponding to an integrated luminosity of 138\fbinv~\cite{CMS-LUM-17-003, CMS-PAS-LUM-17-004, CMS-PAS-LUM-18-002}.
Multijet events are selected using HLT dijet triggers that apply thresholds on the average $\pt$ of the jets having the highest ($p_{\mathrm{T,j_{1}}}$) and the second highest ($p_{\mathrm{T,j_{2}}}$) $\pt$ in an event.
\begin{equation}
        \HTT = \frac{1}{2}\left(p_{\mathrm{T,j_{1}}}+p_{\mathrm{T,j_{2}}}\right).
\end{equation}
The values of the $\HTT$ threshold used in the set of triggers are 60, 80, 140, 200, 260, 320, 400, and 500\GeV. 
These triggers have different prescale factors depending on the trigger rates.
Reconstructed events selected at L1 trigger are used to check if an event passed the appropriate HLT dijet trigger(s) and hence evaluate the efficiency with an emulation method.
The trigger turn-on points, \ie, the values of $\HTT$ at which the trigger is 99\% efficient, are slightly different for each year (2016, 2017, and 2018) of data-taking, as shown in Table~\ref{tab:turnon}, because of the variations in data-taking conditions and HLT. 
This is found not to affect the analysis results.
The trigger turn-on values for the year 2018 are the most conservative and the data sample is divided into eight $\HTT$ ranges accordingly.

\begin{table}[h] 
\centering
\topcaption{The $\HTT$ threshold values, and trigger turn-on points, for the 2016, 2017, and 2018 data sets.}
\label{tab:turnon}
\begin{tabular}{cccc}
\hline     
                  $\HTT$ threshold	& \multicolumn{3}{c}{Turn-on values (\GeVns)}\\
\cline{2-4} 
                               (\GeVns)	& 2016    & 2017      & 2018   \\
\hline 
                           60   &  66      &  84       & 83     \\
                           80  	&  95      &  111      & 109    \\
                          140	&  156     &  172      & 176    \\
                          200	&  227     &  241      & 247    \\
                          260	&  287     &  309      & 318    \\
                          320	&  350     &  376      & 387    \\
                          400	&  444     &  462      & 477    \\
                          500	&  557     &  568      & 573    \\
\hline
\end{tabular}
\end{table}

\subsection{Simulated events\label{sec:sim}}
The \PYTHIAviii v8.240~\cite{Sjostrand:2014zea}, \MGvATNLO(MG5) v2.6.5~\cite{Alwall:2014hca}, and \HERWIG{}\,7 v7.2.2~\cite{Bellm:2015jjp} generators are used to simulate $\Pp\Pp$ collision events at $\sqrt{s} = 13\TeV$. 
The NNPDF3.1 NNLO~\cite{NNPDF:2017mvq} parton distribution function (PDF) set is used along with the leading-order matrix elements. 
This provides a good description of a broad range of LHC measurements~\cite{Sirunyan:2019dfx}. 
It is found that the PDF uncertainties in the normalized unfolded distributions of the ESVs are small, and hence the order of the PDF set is not expected to have a significant effect.
Simulated events are passed through the CMS detector response and reconstruction package based on \GEANTfour~\cite{Agostinelli:2002hh}. The simulated events are selected using the same criteria as used for the data.

The \PYTHIAviii generator operates at leading order (LO), and uses $2\to2$ matrix element (ME) calculations at the Born level, followed by a $\pt$-ordered PS~\cite{Sjostrand:2004ef, Corke:2010yf}, which includes initial- and final-state radiations (ISR and FSR). Multi-parton interactions (MPIs) and hadronization based on the Lund string model~\cite{Andersson:1983ia} complete the generation of the event. The CP5 tune~\cite{Sirunyan:2019dfx} of \PYTHIAviii is used in this study to generate the multijet sample.

The \HERWIG{}\,7  generator also produces LO events, but uses an angular-ordered~\cite{Banfi:2006gy} PS along with the colour coherence effects~\cite{Gieseke:2012ft} and cluster model~\cite{Gieseke:2016fpz} for the hadronization. This study uses the CH3~\cite{CMS:2020dqt} tune of \HERWIG{}\,7 to generate the multijet sample.

The \mgpy generator uses tree-level ME calculations for the hard-scattering events. In this study, \MGvATNLO generates $2\to2$, $2\to3$, and $2\to4$ events at LO followed by PS and hadronization by the \PYTHIAviii CP5 tune. Jet matching is performed using the MLM~\cite{Hoche:2006ph} prescription to avoid the double counting of jets between the ME and PS.

\subsection{Event selection\label{sec:evntsel}}
The jets are required to satisfy selection criteria~\cite{CMS:2017wyc,CMS:2013pia} to reject spurious jets from detector noise and pileup. 
At both collision and particle level, for an event to be selected, it must have ${\geq}2$ selected jets with $\pt > 30\GeV$ and the two highest \pt jets in the event must be within $\abs{\eta} \le 2.1$.
We use all the charged particles inside the jets ($\abs{\eta_{\text{jet}}} \le 2.1$) where the charged particles are within $\abs{\eta} < 2.5$ to evaluate the ESVs.
At particle level, a mean decay length ~$c\tau > 10\mm$ is used to define stable particles.

Track reconstruction imposes a $\pt$ threshold of $> 0.5$\GeV~\cite{CMS:2017yfk}. We have checked the effect of lowering the $\pt$ threshold in the MC and observed that the effect is largest in the lowest $\HTT$ range. The mean of the ESV distributions vary by $<$1\%, while the bin-wise variations are mostly within 1--2\% and reach up to 5\% in a few bins and the shapes of the distributions remain mostly unchanged.

\section{Unfolding of distributions\label{sec:unf}}
Detector-related factors, \eg, incomplete coverage, finite resolution, particle identification efficiency, misidentification, and detector noise, affect the reconstruction of an event and hence the evaluation of the ESVs. The distribution of an ESV obtained from data (reconstructed level) needs to be corrected for the detector effects to obtain the true (particle level) distribution.
This is required for the comparison of the distributions with the theoretical predictions and results from other experiments.

The possible effects of the detector on this study are as follows. 
An event selected at the particle level may be rejected at the reconstructed level.
Also, an event belonging to one $\HTT$ range at the particle level may migrate to another after reconstruction. 
An event that is not selected at the particle level may be selected at the reconstructed level. 
Furthermore, the value of an ESV at the reconstruction level may vary from that at the particle level.

For each ESV a two-dimensional (2D) unfolding involving the variable and $\HTT$ is performed using the \tunfold~\cite{Schmitt:2012kp} framework to obtain the underlying particle-level distributions from the reconstructed ones.
Events generated with an MC event generator are passed through a detailed detector simulation, that provides all the required information to construct the response matrices (RMs) and 2D unfolding.
Events generated with \PYTHIAviii~(CP5 tune) are used as the default for unfolding throughout this study. 

The resolution of an ESV is taken into account when determining the bin widths for its distribution, to ensure that purity is ${\ge}50\%$ for each bin.
Purity is determined as the fraction of the events in a distinct reconstructed level bin that are generated from the same bin at the generator level.
This ensures stable unfolding of the variables, and no regularization~\cite{Schmitt:2016orm} is required.
Misidentified events are subtracted from the reconstruction-level distributions, and the unfolded distributions are corrected for inefficiency to obtain the particle-level distributions. A number of closure tests have been performed which show that the unfolding procedure is correct and accurate.
The systematic uncertainties associated with the RM are also taken into account in the unfolding. 

\section{Uncertainties\label{sec:unc}}
The major sources of uncertainty in the particle-level distributions of the ESVs are discussed below.
Contributions from different sources are added in quadrature to obtain the total uncertainty in each bin of the unfolded distribution of an observable.
A summary of the contributions of different sources of uncertainty is given in Table~\ref{tab:unc}.

\subsection{Experimental systematic uncertainties}
\textit{Jet energy scale (JES)}: A significant contribution to the systematic uncertainty arises from the JES. 
CMS uses a factorized approach to calibrate the energy of each jet, which includes corrections due to pileup and detector or reconstruction effects~\cite{Khachatryan:2016kdb}.
Altogether, CMS takes into account 27 different sources of systematic uncertainty for the JES in Run~2~\cite{CMS-DP-2020-019}.
For each individual source, the four-momenta of each jet are raised or lowered by the respective uncertainty and this is propagated through the unfolding procedure to obtain the observable with the JES shifted up or down.
In each bin, the larger of the differences between the nominal and the two varied predictions is taken as the systematic uncertainty for that source. 
Uncertainties from all the sources are added in quadrature to obtain the total uncertainty due to the JES for that bin. The contribution of the JES to the relative uncertainty for the unfolded distributions of the ESVs is ${\sim}1$--$3\%$, as shown in Table~\ref{tab:unc}.  

\textit{Jet energy resolution (JER)}: It has been observed that the JER, which is a function of both $\eta$ and $\phi$ of the jet, is worse in data than in the simulated events, and additional smearing has to be applied to jets in the latter~\cite{Khachatryan:2016kdb}.
The degree of additional smearing is varied up and down, the corresponding RMs are constructed, and the detector-level distributions of an ESV are unfolded.
For each bin, the larger of the differences with the nominal one is taken as the associated uncertainty. 
The contribution of JER to the relative uncertainty in the unfolded distributions of the ESVs is ${\approx}1\%$.

\textit{Pileup}: Uncertainty due to pileup subtraction is estimated following the same procedure as for the JER.
For calculating the uncertainty due to pileup, the total inelastic cross section~\cite{CMS:2020ebo} has been varied by its uncertainty of 4.6\%.
The MC samples are weighted with pileup profiles calculated with scaled-up and scaled-down cross sections, the corresponding RMs are constructed, and the uncertainties determined from the unfolded results.
The uncertainty due to pileup is 1--2\%.

\textit{Track reconstruction efficiency}:
The uncertainty due to track reconstruction efficiency is estimated using the pion reconstruction results in $\PDst$ decays for 2016--2018 data~\cite{CMS:2010mua,CMS-DP-2022-012}. 
Tracks are randomly removed in events at the reconstruction level in simulated events, RMs are formed, and the unfolding is repeated.
The difference from the nominal result is taken as the uncertainty.
The relative uncertainty due to track reconstruction inefficiency is ${\approx}1\%$.

\textit{Track $\pt$ resolution}:
The track \pt resolution varies with the $\pt$ and $\eta$ of the track~\cite{CMS:2014pgm}. For each charged particle, the $\pt$ is varied within the resolution uncertainties and the RMs are constructed for the \PYTHIAviii event sample to unfold the data.
The difference due to the variation in track $\pt$ is taken as the uncertainty.
The uncertainties due to track \pt resolution are found to be small, $\mathrm{<1\%}$.

\textit{ECAL L1 prefiring}: Data-taking during 2016 and 2017 was affected by a gradual shift in the timing of the ECAL L1 trigger primitives.
As a result, in the region $\abs{\eta}> 2.0$, a trigger signal was mistakenly associated with the previous bunch crossing (known as ``prefiring")~\cite{CMS:2024ppo}.
This inefficiency is corrected by using an event weight associated with the nonprefiring probability computed from data and applied to the simulation.
The event weight is varied by its uncertainty to estimate the uncertainty due to trigger prefiring.
The contribution of the ECAL L1 prefiring to the relative uncertainty is small, $<$1\%.

\subsection{Theoretical systematic uncertainties}
\textit{Monte Carlo model dependence}: Simulated events passed through the detector simulation are used to construct RMs, introducing a dependence of the unfolded distribution on the chosen event generator and its QCD model. 
Theoretical uncertainties due to model dependence are estimated by unfolding detector-level distributions using RMs from \PYTHIAviii, \mgpy, and \HERWIG{}\,7.
The uncertainties associated with misidentification of events and inefficiency corrections are found to be small, typically below 1\%, and are included in the total MC modelling uncertainty.
The uncertainty is typically larger for the bins towards the edges of a distribution, where the number of entries is small. Also, there are some occasional fluctuations and polynomial interpolation~\cite{Connor:2021tzg} is used for smoothing. Overall, this uncertainty is typically 1--5\% and decreases with increasing $\HTT$.

\textit{ISR, FSR scale}: The uncertainty due to the renormalization scale in the ISR and FSR within \PYTHIAviii~\cite{Mrenna:2016sih} is estimated. The contribution due to ISR is found to be small, $\mathrm{<1\%}$ and the combined uncertainty due to the scale variation is  1--2\%.

\textit{$\mu_R$, $\mu_F$ scale}: The uncertainty due to the renormalization and  factorization scale in the matrix element with \mgpy  is estimated. 
The uncertainty is found to be $<$1\% for $\taup$, and for the four other variables, it is typically 1\%, reaching up to about 2--4\% in a 
few bins of the lower $\HTT$ range.

\textit{Parton distribution function (PDF)}:
The uncertainty due to PDFs is estimated using 100 sets of replica PDFs that reflect the PDF uncertainties provided by the NNPDF3.1 set.
The standard deviation of these values for a bin is taken as the uncertainty for that bin due to the choice of PDF.
This contribution is small, $<$1\%, for all the five variables. 
\subsection{Statistical uncertainty}
The uncertainty due to the finite number of data events used in this analysis has been estimated. 
Similarly, the uncertainty due to the finite number of MC events to estimate the efficiencies, RMs, and misidentified event contamination has been estimated. 
The contributions of both are small and the combined uncertainty is $<$1\% for all the ESVs, see Table~\ref{tab:unc}. 

\subsection{Total uncertainty}
Uncertainties from different sources are added in quadrature to obtain the total uncertainty in the unfolded distributions.
The total relative uncertainty is 1--4\% for $\taup$, while it may go up to 6\% in some $\HTT$ ranges for the other variables, as shown in Table~\ref{tab:unc}.
The contributions from the JES correction and the MC model dependence dominate the relative uncertainty in the unfolded distributions of the ESVs, and these reduce with increasing $\HTT$ values.
\begin{table}[h]
\centering
\topcaption{Typical values of the relative uncertainties in the normalized unfolded distributions of the five ESVs.}
\label{tab:unc}
\begin{tabular}{lccccc}
\hline
Source of uncertainty	& \multicolumn{5}{c}{Relative uncertainty (\%)}\\
\cline{2-6}
                        & $\taup$ &  $\mass$ & $\ytwo$ & $\bT$ & $\massp$ \\
\hline
Jet energy scale	& 1--2 	& 1--2	& 1--2	& 1--3	& 1--3 \\
Jet energy resolution 	&  $<$1 &  $<$1 & $<$1 & $<$1  &  $<$1 \\
Pileup 	 	& ${\approx}$1  & 1--2   & 1--2  & 1--2  & 1--2  \\
Unfolding 		& $<$1   & $<$1   & $<$1  & $<$1  & $<$1 \\
Track reconstruction efficiency 
                        & ${\approx}1$  & 1--2  & 1--4  & ${\approx}1$ & ${\approx}1$ \\
Track $\pt$ resolution 	&  $<$1  & $<$1 & $<$1 & $<$1  & $<$1 \\
ECAL L1 prefiring 	& $<$1  & $<$1  & $<$1  & $<$1  & $<$1  \\[\cmsTabSkip]

        {Experimental total} & 1--2	& 1--3 & 1--4 & 1--3 & 1--4  \\[\cmsTabSkip]

Monte Carlo model dependence	
                        & 1--4  & 2--4  & 2--5  & 2--4  & 1--5 \\
PDF	& $<$1 & $<$1  & $<$1 & $<$1  & $<$1  \\[\cmsTabSkip]
ISR, FSR scale	& $<$1 & 1--2  & 1--2 & 1--2  & 1--2  \\[\cmsTabSkip]
$\mu_{R}$, $\mu_{F}$ scale	& $<$1 & 1--2  & 1--3 & 1--4  & 1--2  \\[\cmsTabSkip]

{Theory (total)}	&1--4 & 2--5 & 2--7 & 2--5 & 2--6\\[\cmsTabSkip]						

{Statistical} 		& ${\approx}1$  & ${\approx}1$ & ${\approx}1$  & ${\approx}1$ & ${\approx}1$  \\[\cmsTabSkip]

{Total uncertainty}	& 1--4	& 2--6	& 2--8	& 2--7	& 2--7	\\
\hline
\end{tabular}
\end{table}

\section{Results\label{sec:result}}
\begin{figure}[!htb]
\centering
\includegraphics[width=0.32\textwidth]{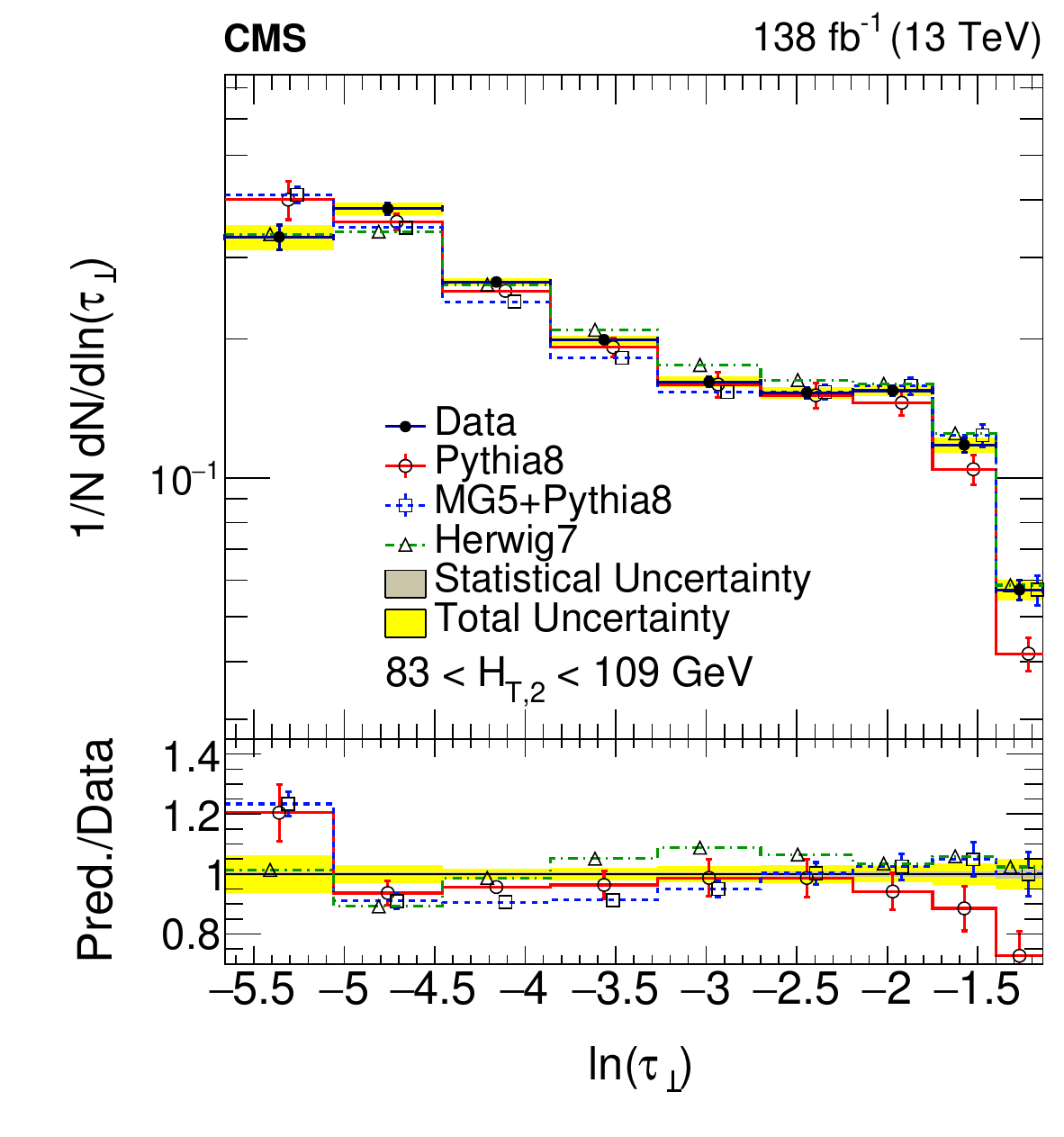}
\includegraphics[width=0.32\textwidth]{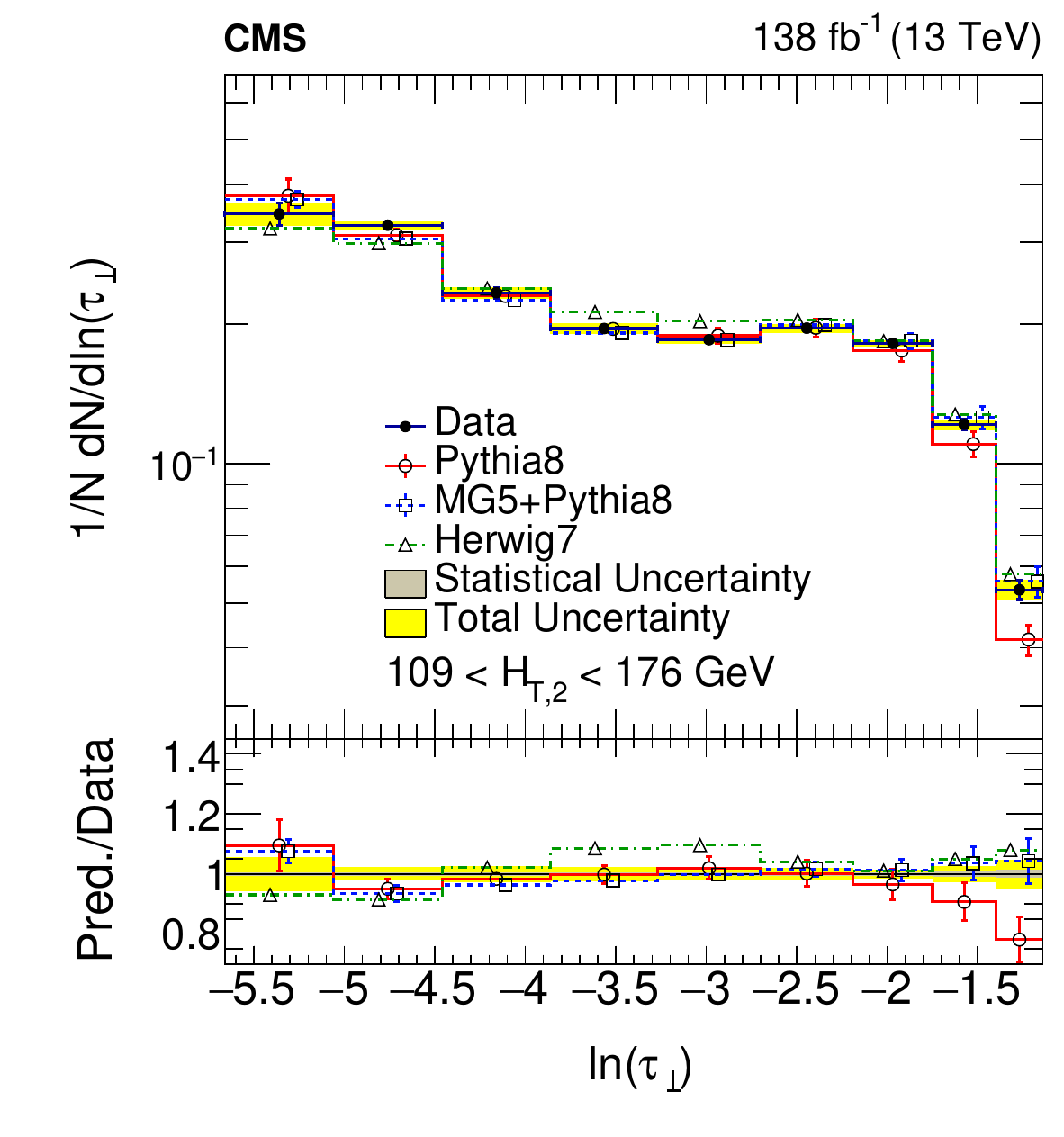}
\includegraphics[width=0.32\textwidth]{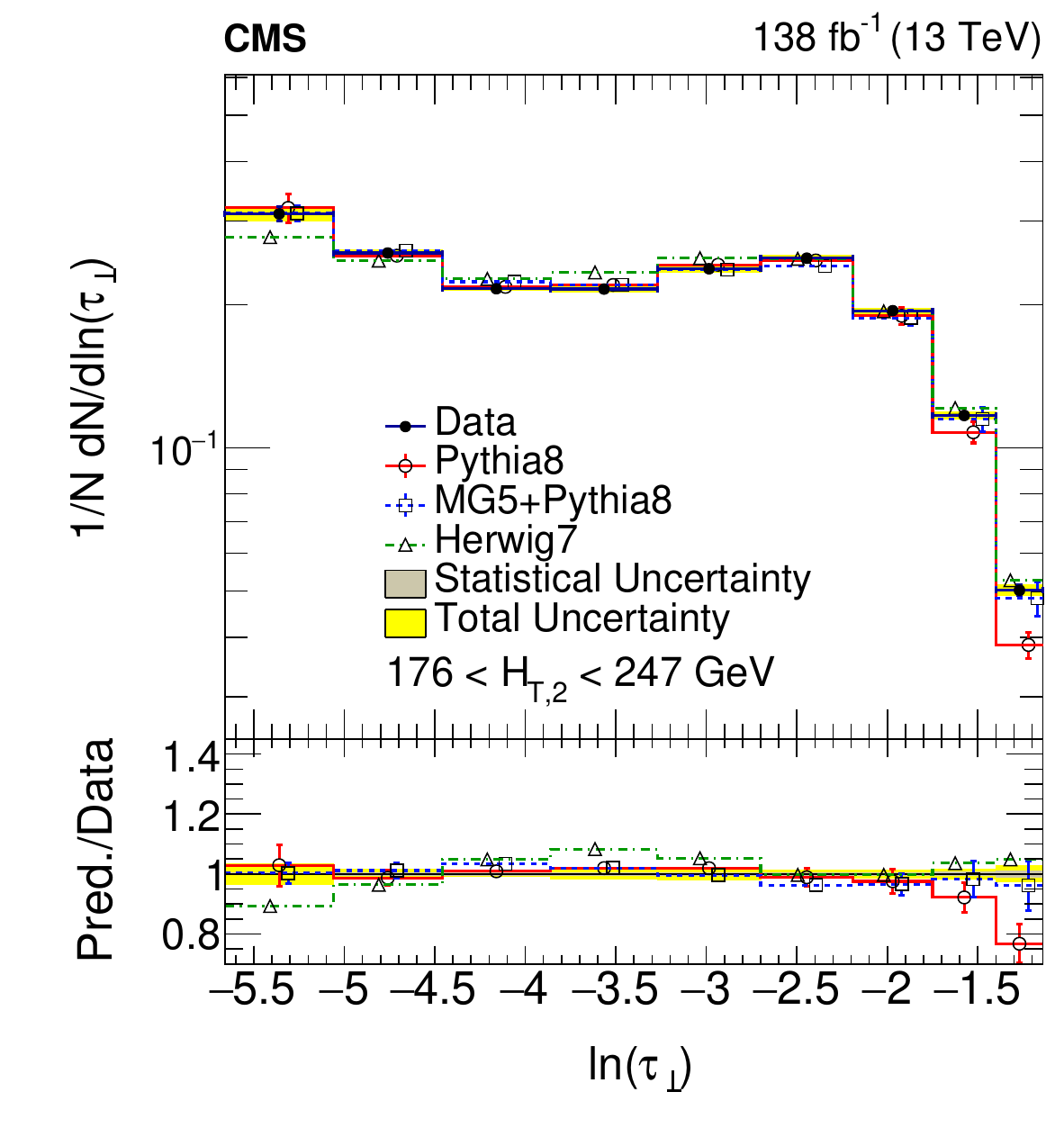}\\
\includegraphics[width=0.32\textwidth]{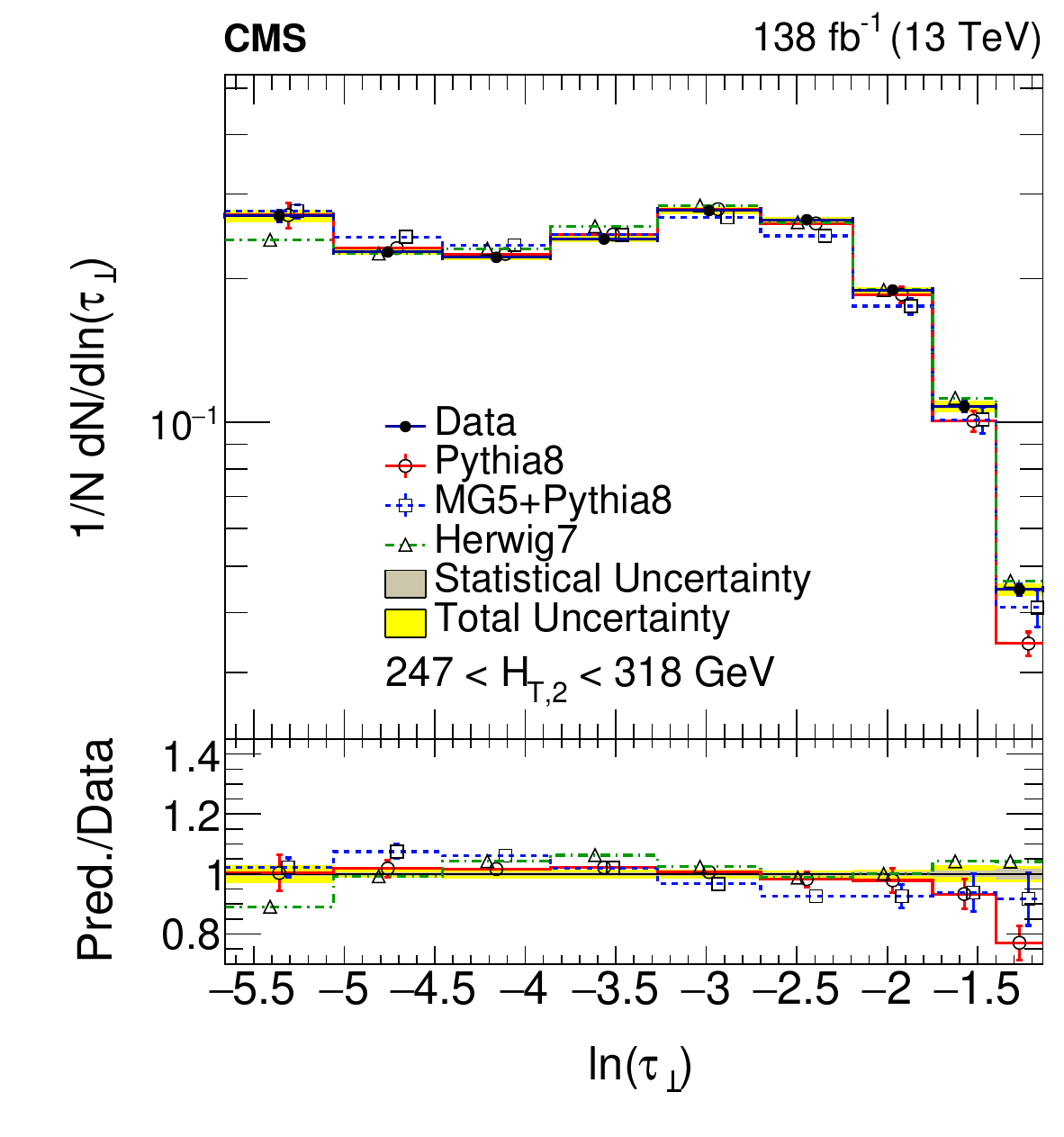}
\includegraphics[width=0.32\textwidth]{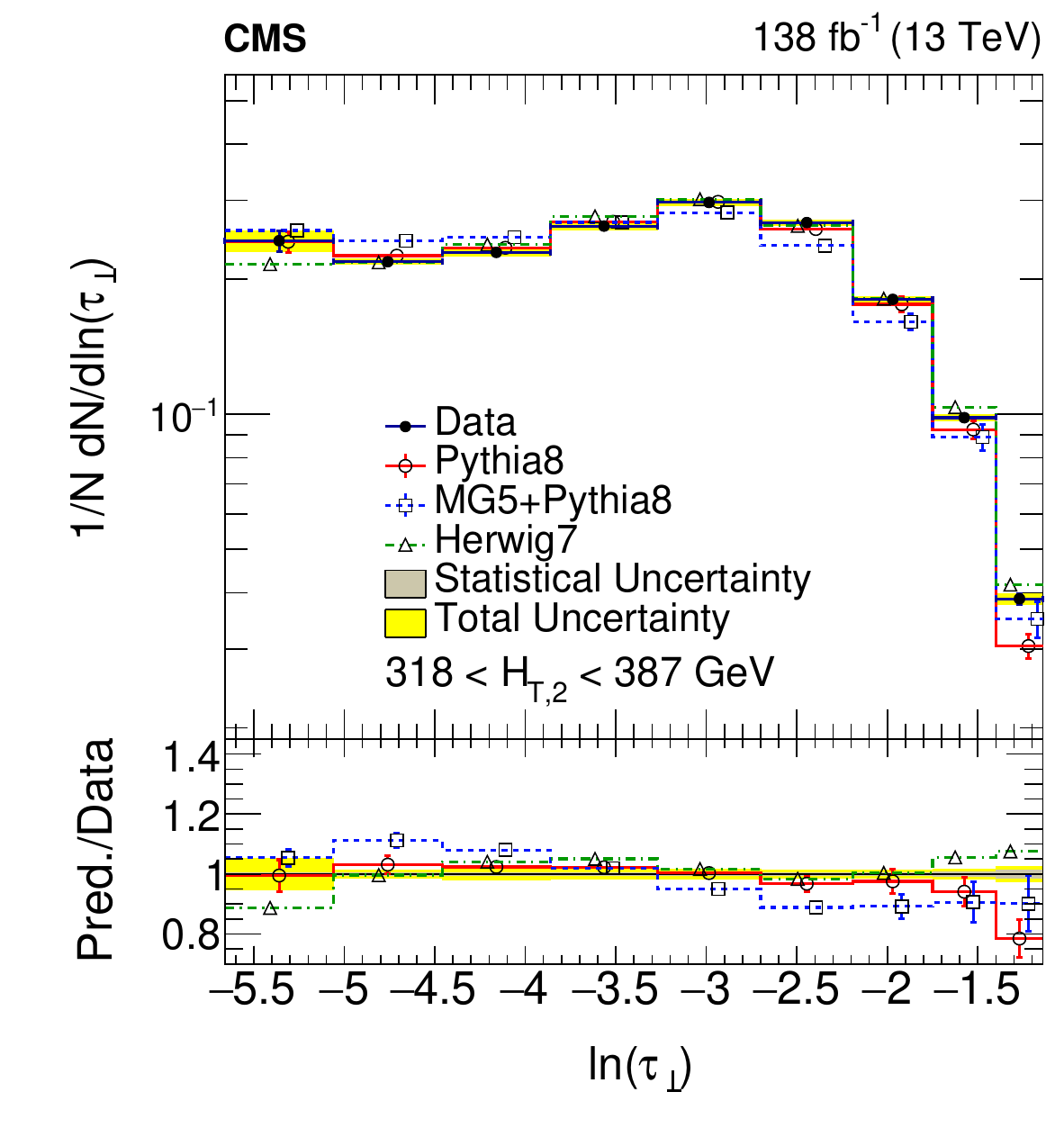}
\includegraphics[width=0.32\textwidth]{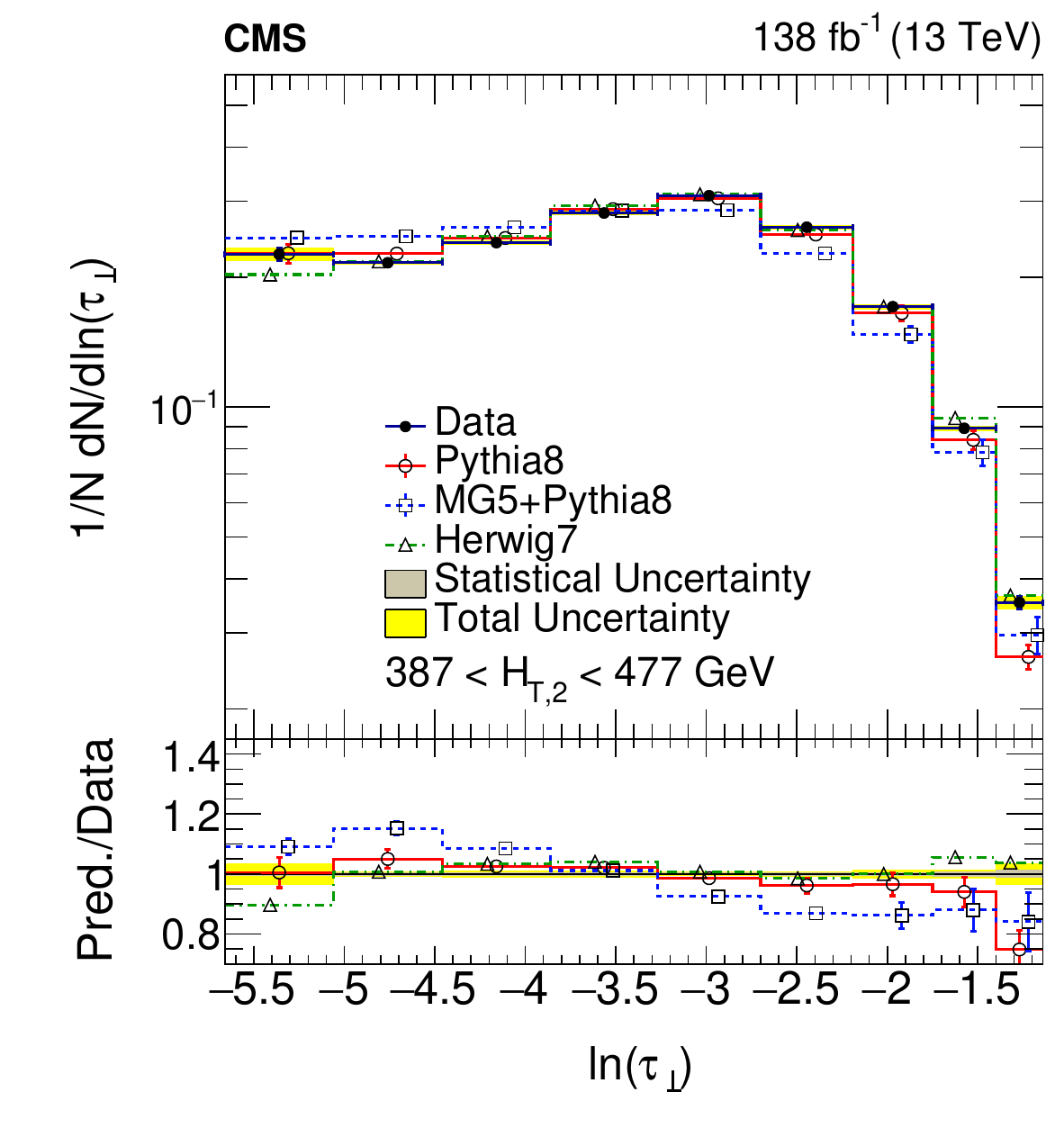}\\
\includegraphics[width=0.32\textwidth]{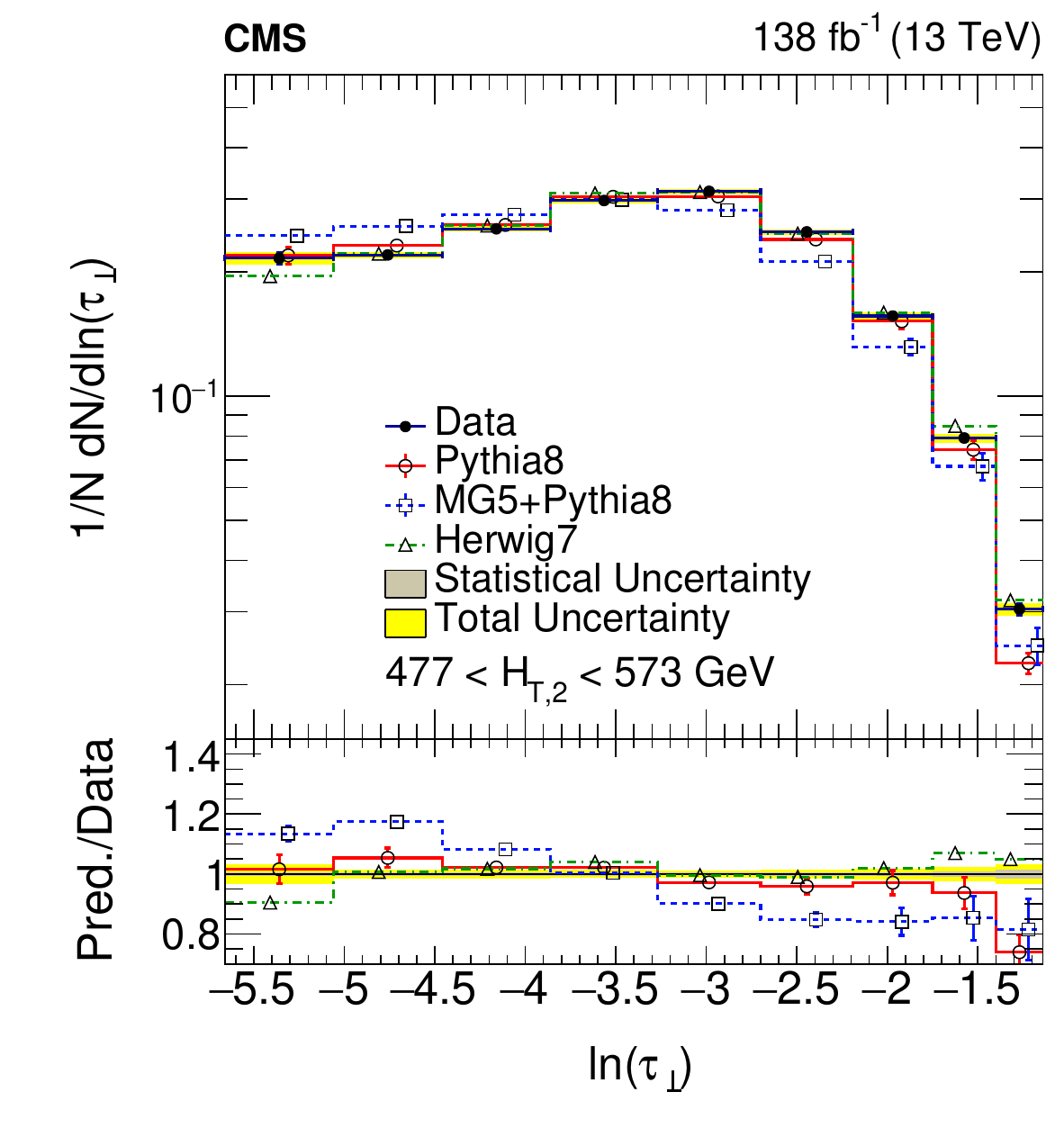}
\includegraphics[width=0.32\textwidth]{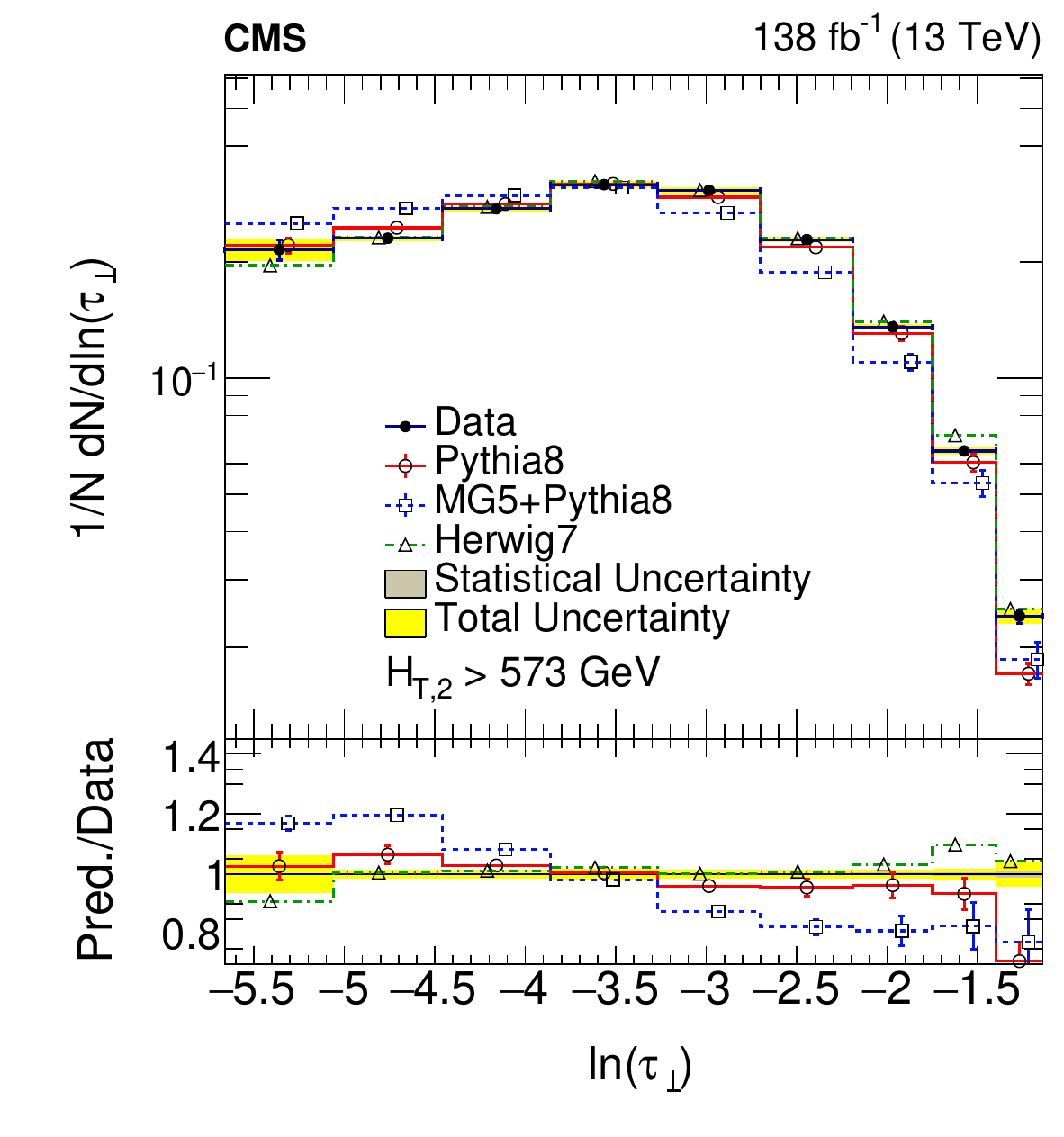}
\caption{Normalized unfolded distributions of the complement of transverse thrust ($\taup$) compared with the predictions of \PYTHIAviii (CP5 tune) (red, solid), \HERWIG{}\,7 (CH3 tune) (green, dash-dotted) and \mgpy (blue, dotted) for different $\HTT$ ranges. For each plot the inner grey band and the yellow band represent the statistical and total uncertainties and the lower panel shows the ratio of simulation to data. The error bars corresponding to \PYTHIAviii (in red) represent renormalization scale variation in ISR and FSR. The error bars corresponding to the \mgpy (in blue) predictions represent the uncertainties due to the renormalization and factorization scale variations.}

\label{fig:thrustRun2}
\end{figure}

\begin{figure}[!htb]
\centering
\includegraphics[width=0.32\textwidth]{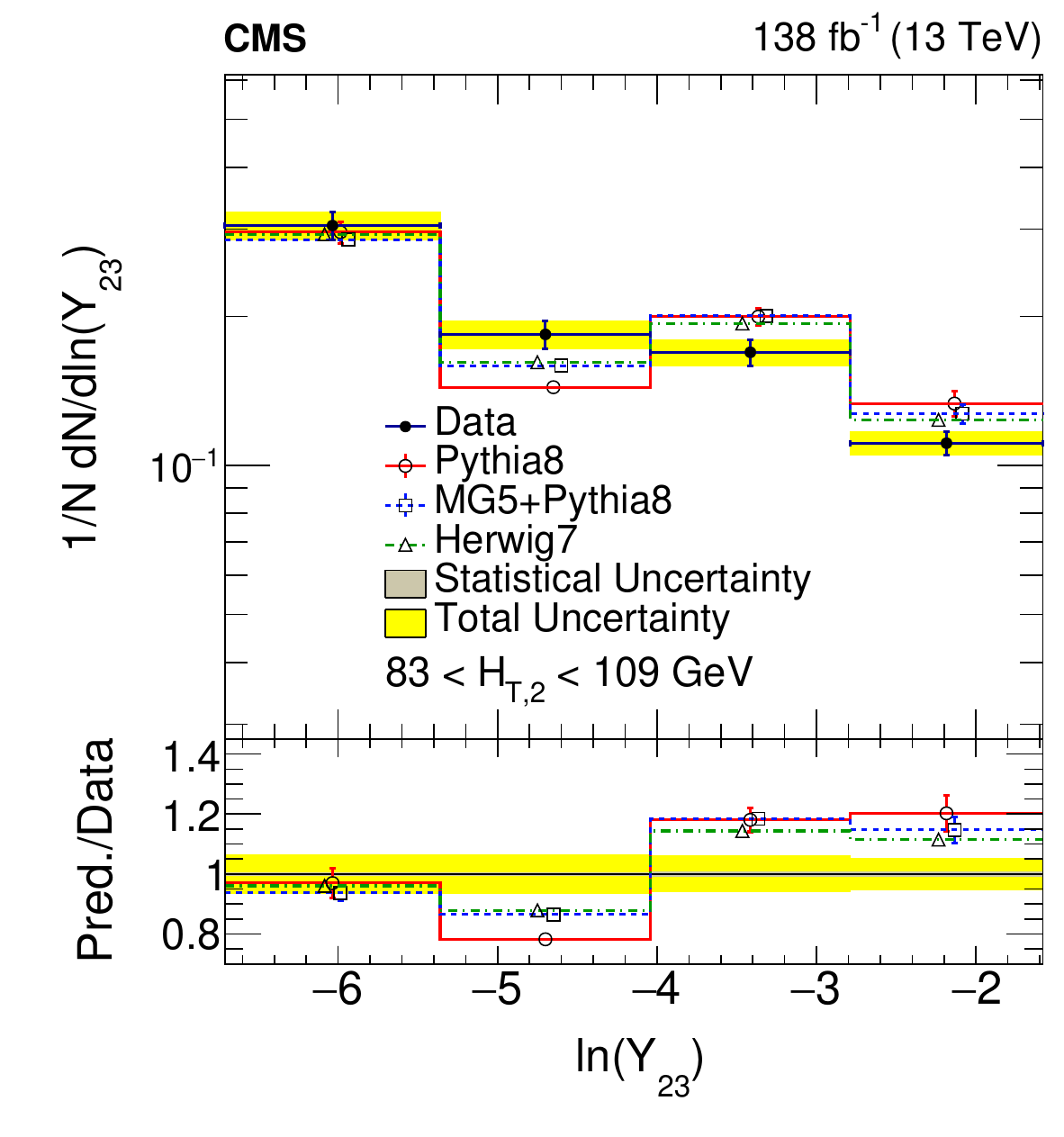}
\includegraphics[width=0.32\textwidth]{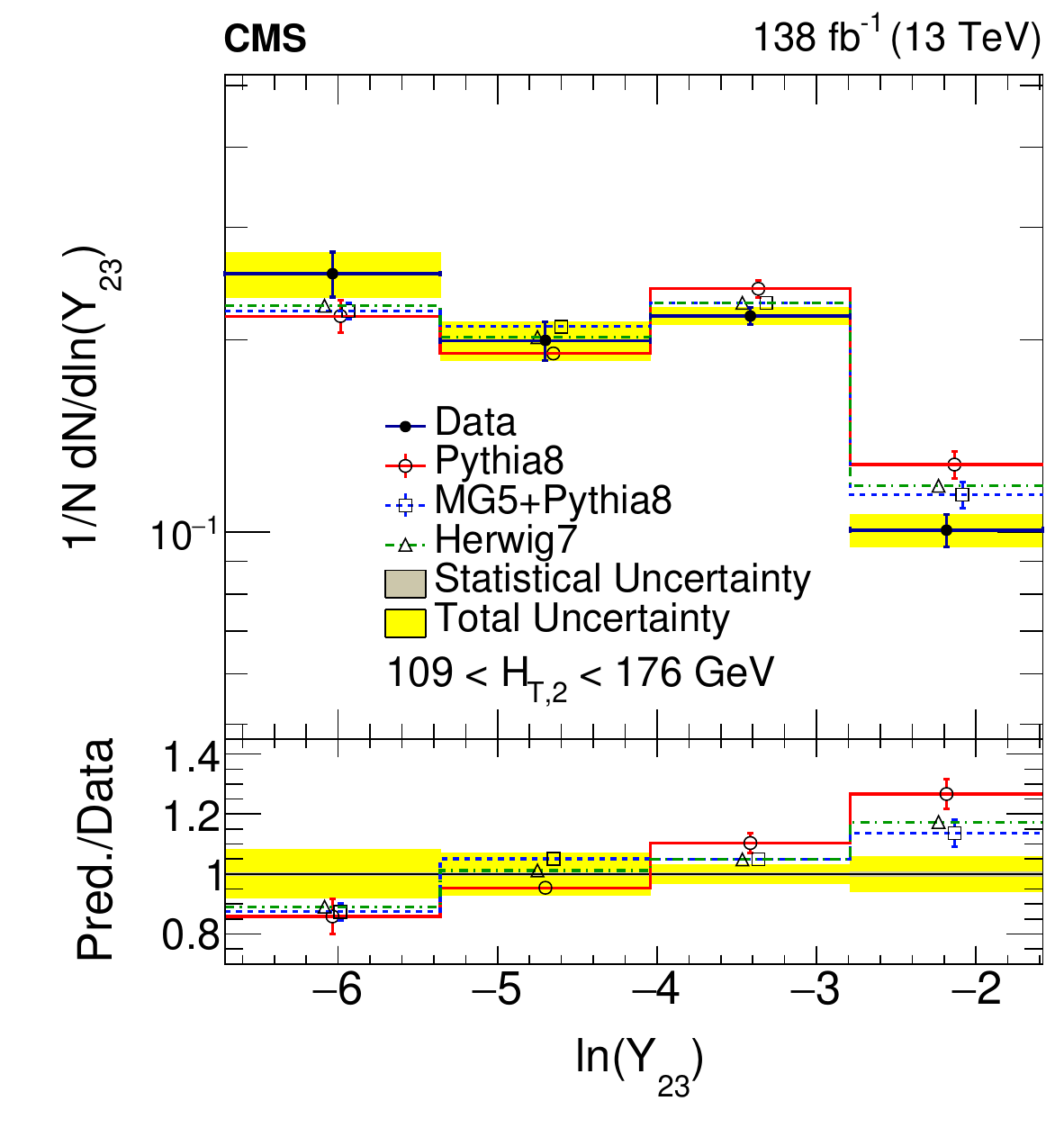}
\includegraphics[width=0.32\textwidth]{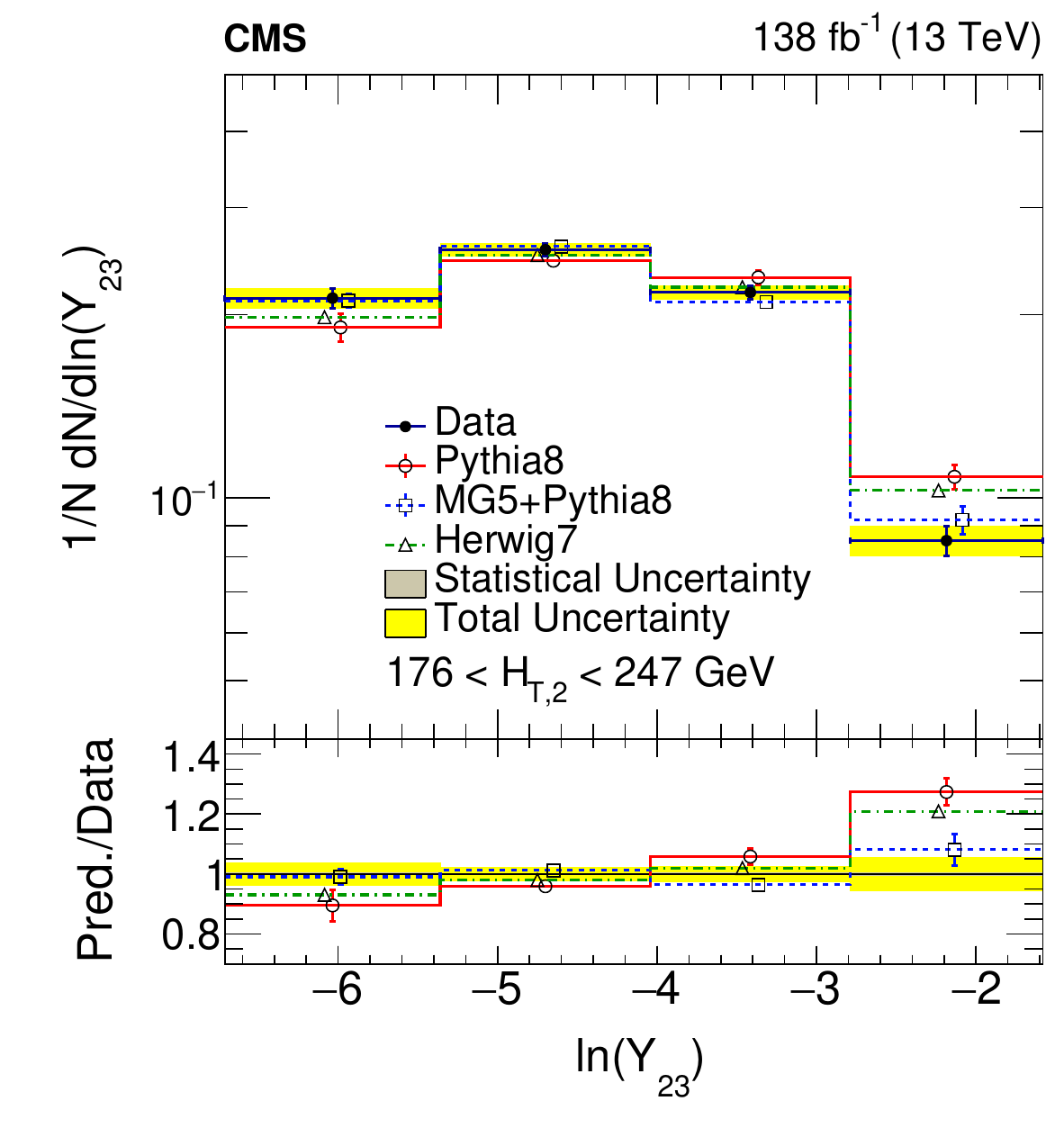}\\
\includegraphics[width=0.32\textwidth]{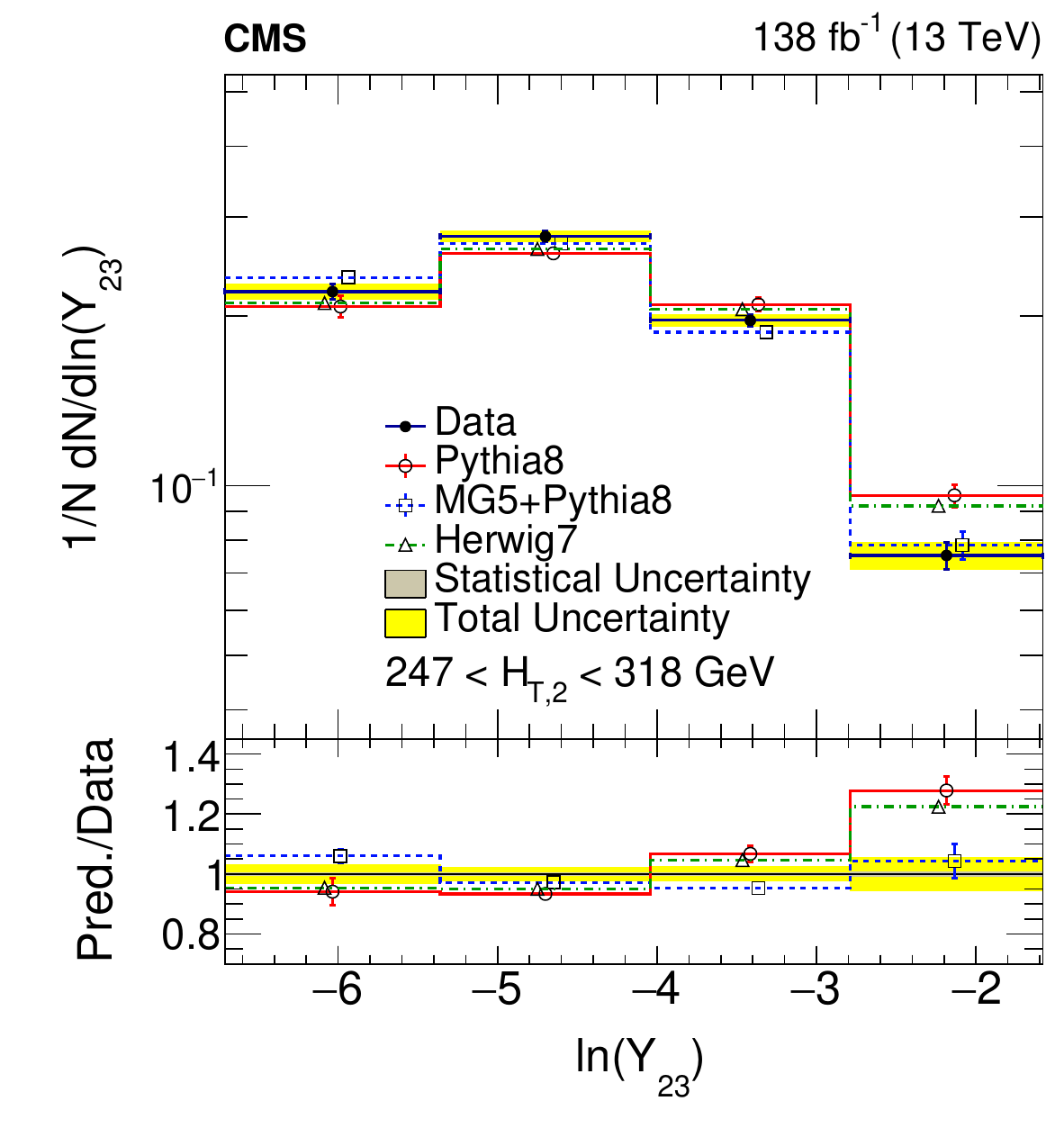}
\includegraphics[width=0.32\textwidth]{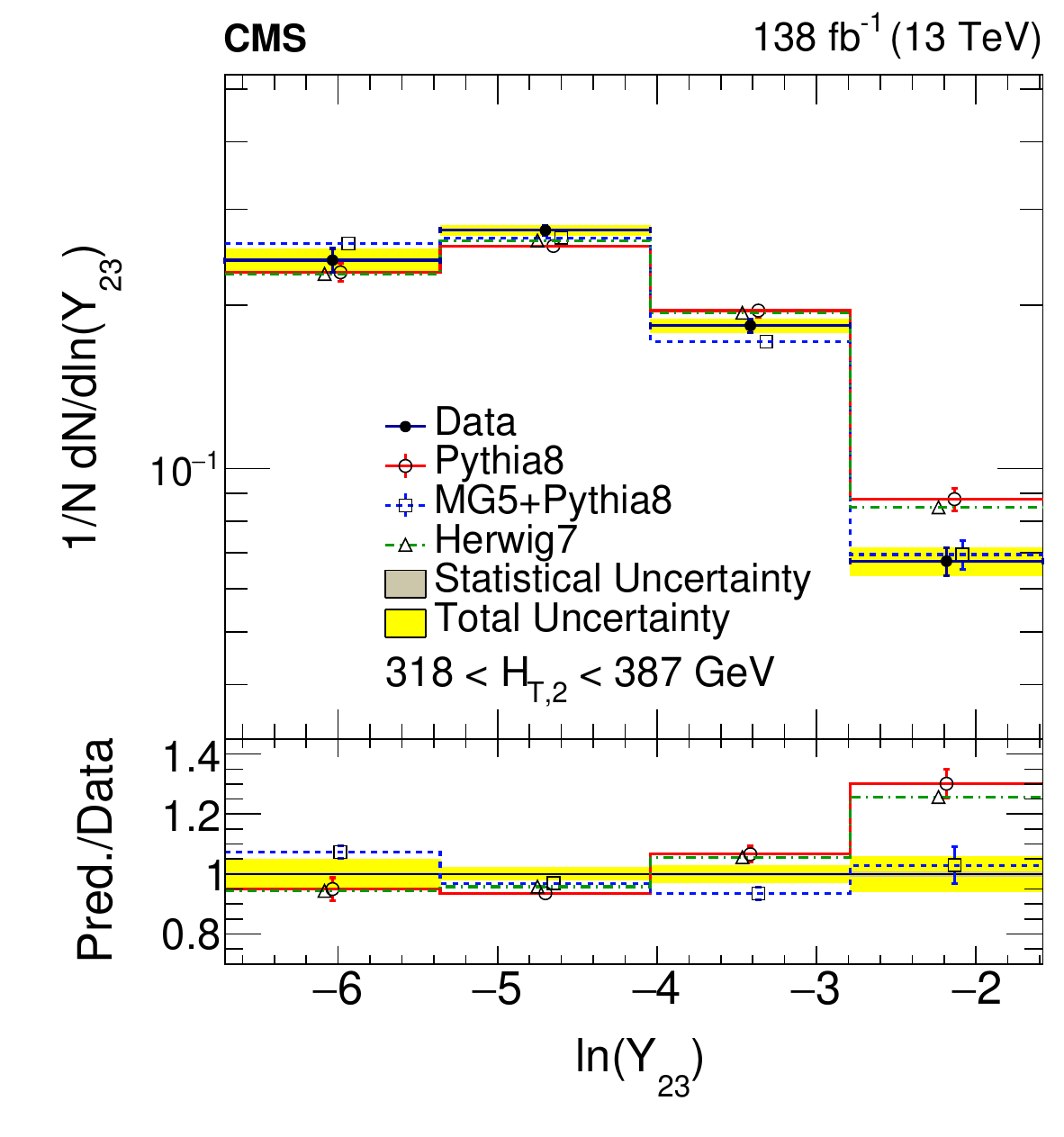}
\includegraphics[width=0.32\textwidth]{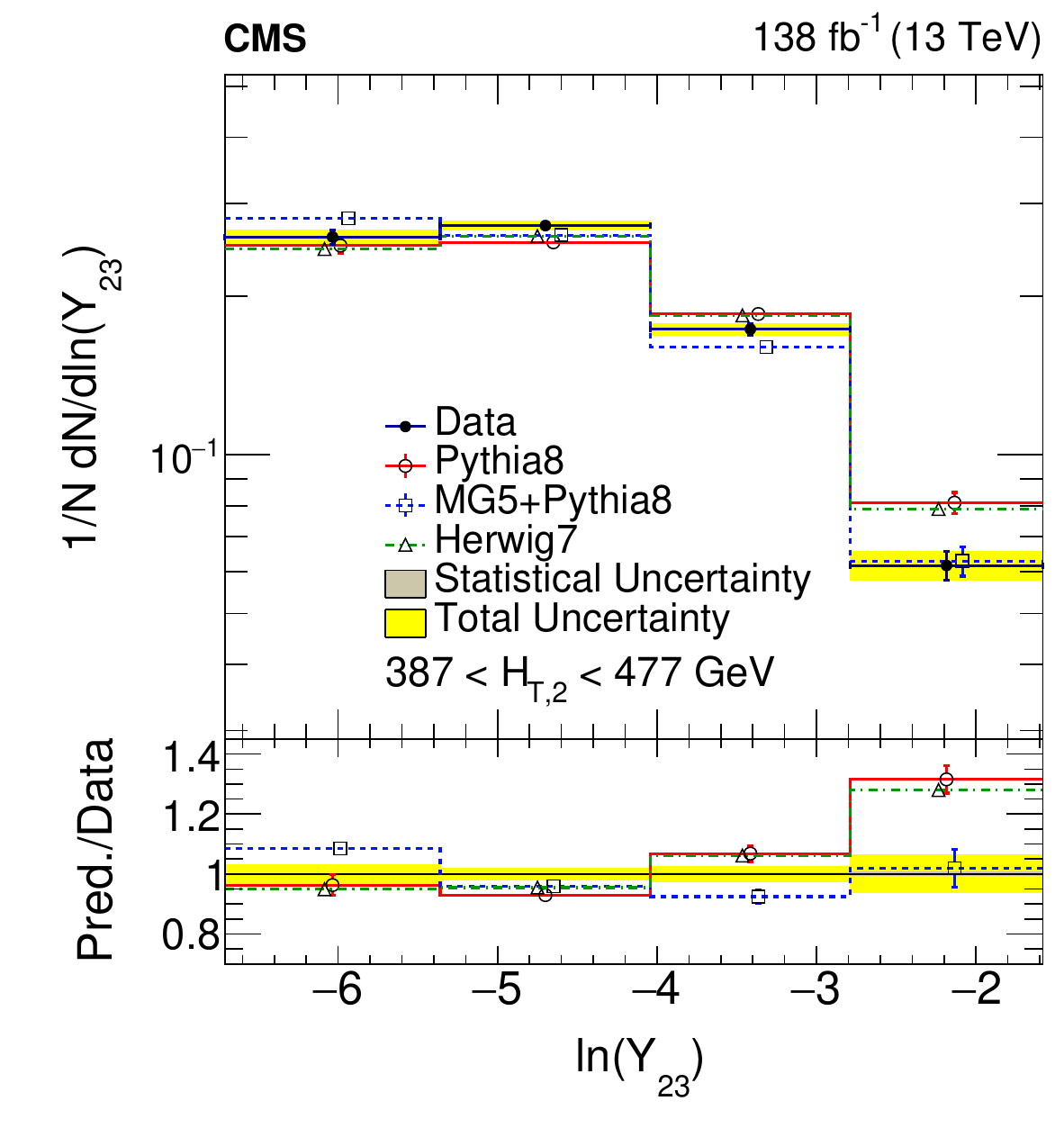}\\
\includegraphics[width=0.32\textwidth]{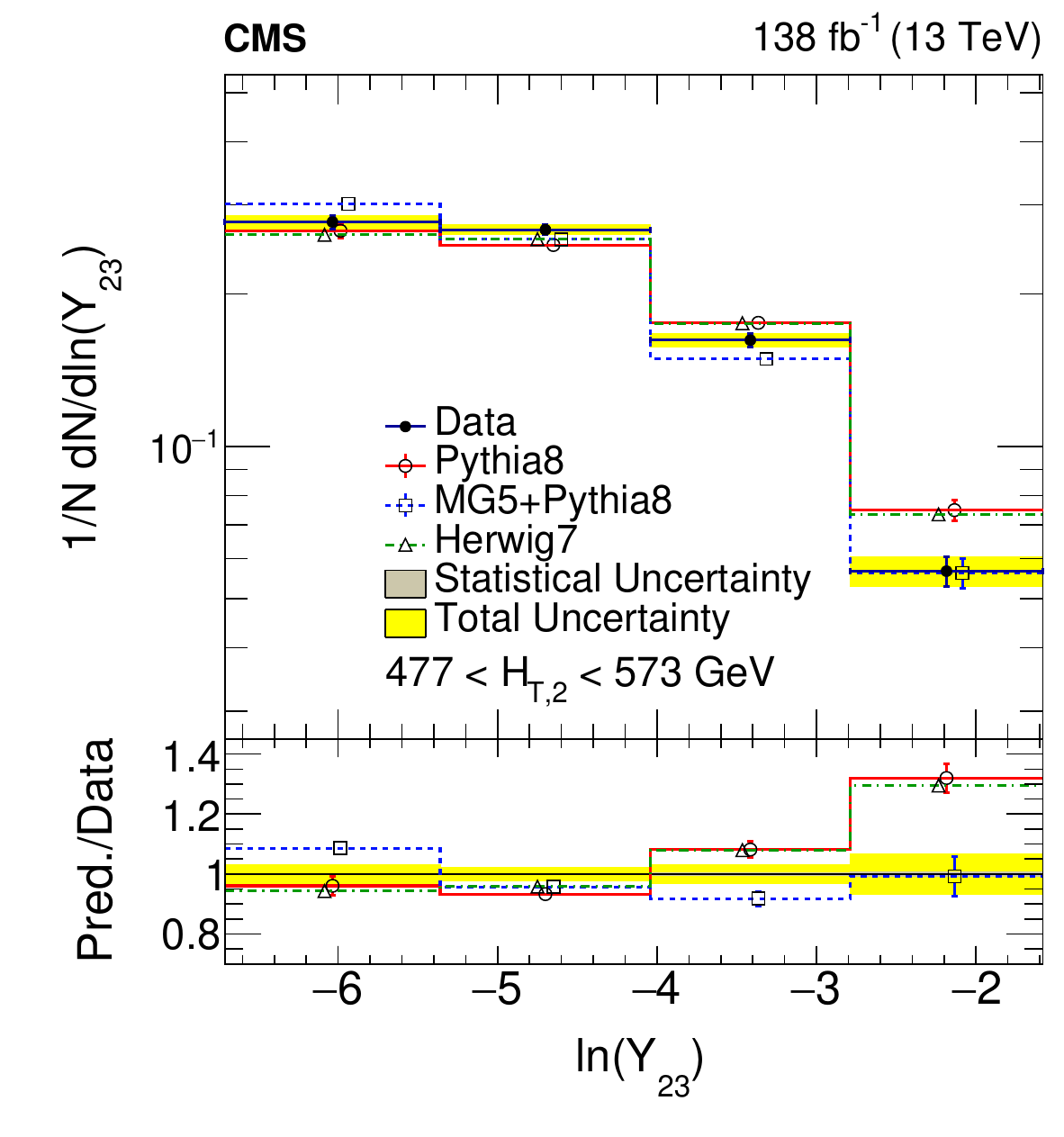}
\includegraphics[width=0.32\textwidth]{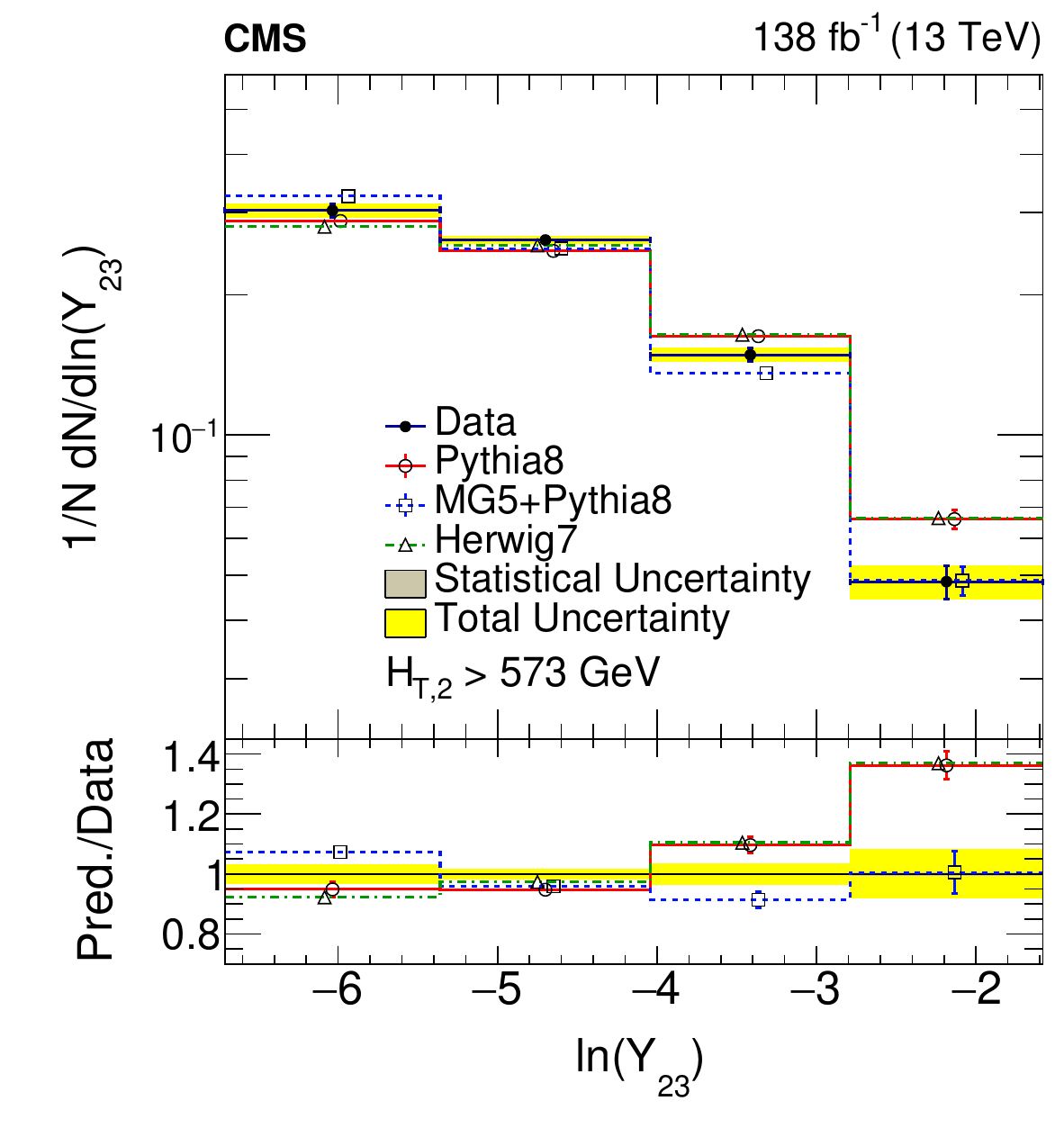}
\caption{Normalized unfolded distributions of the third jet resolution parameter ($\ytwo$) compared with the predictions of \PYTHIAviii (CP5 tune) (red, solid), \HERWIG{}\,7 (CH3 tune) (green, dash-dotted) and \mgpy (blue, dotted) for different $\HTT$ ranges. Notations as in Fig.~\ref{fig:thrustRun2}.}
\label{fig:y23Run2}
\end{figure}

\begin{figure}[!htb]
\centering
\includegraphics[width=0.32\textwidth]{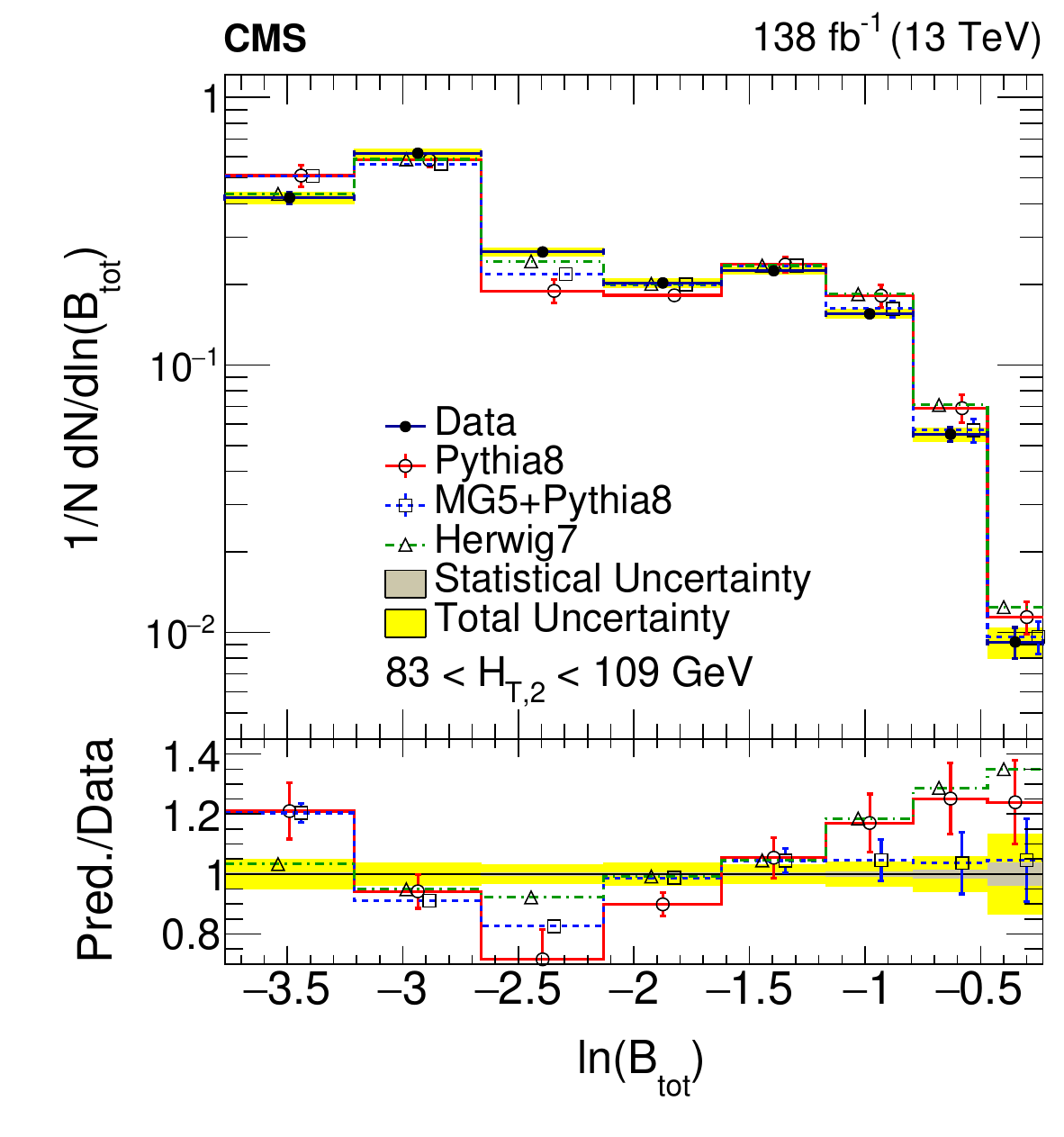}
\includegraphics[width=0.32\textwidth]{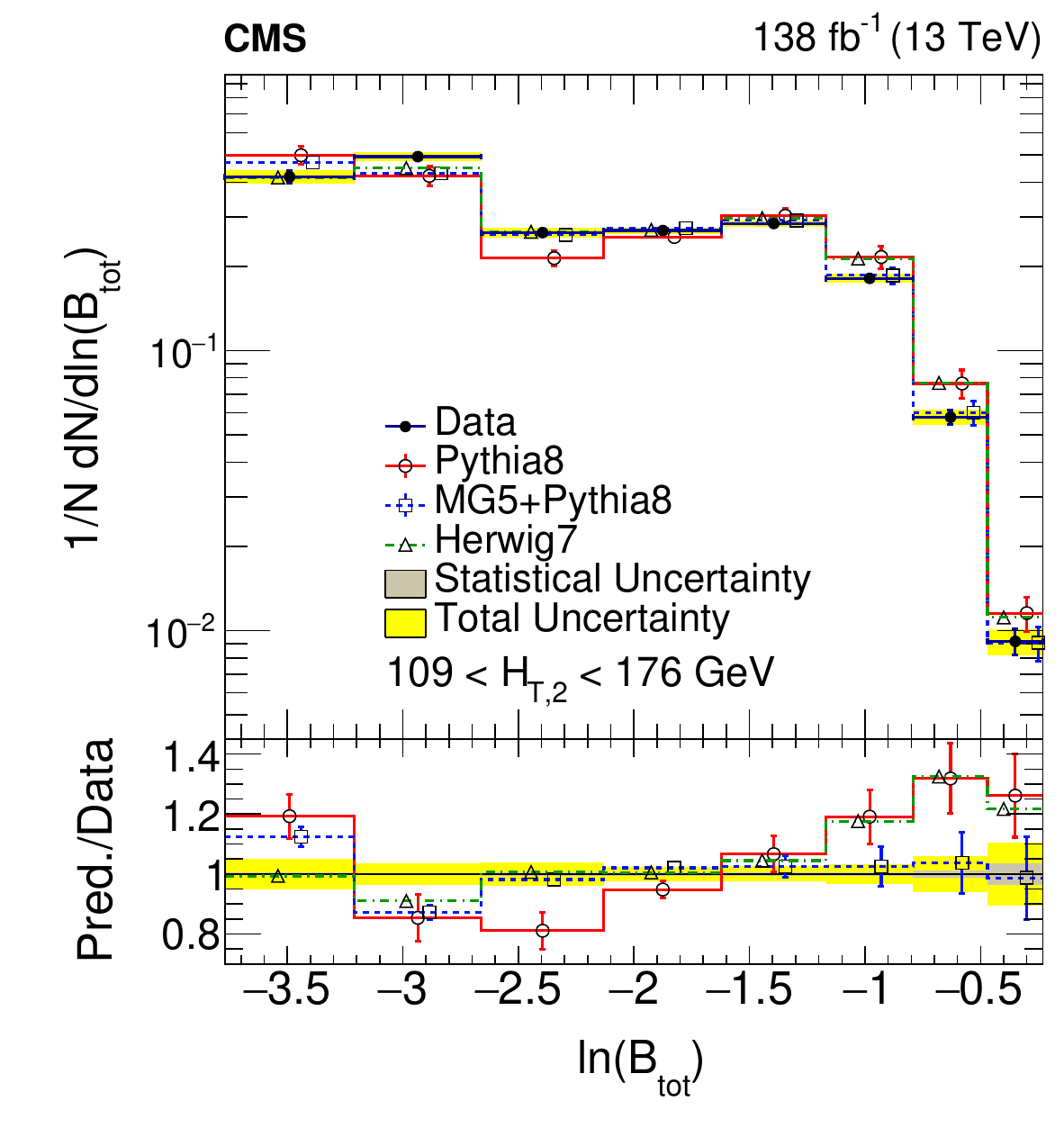}
\includegraphics[width=0.32\textwidth]{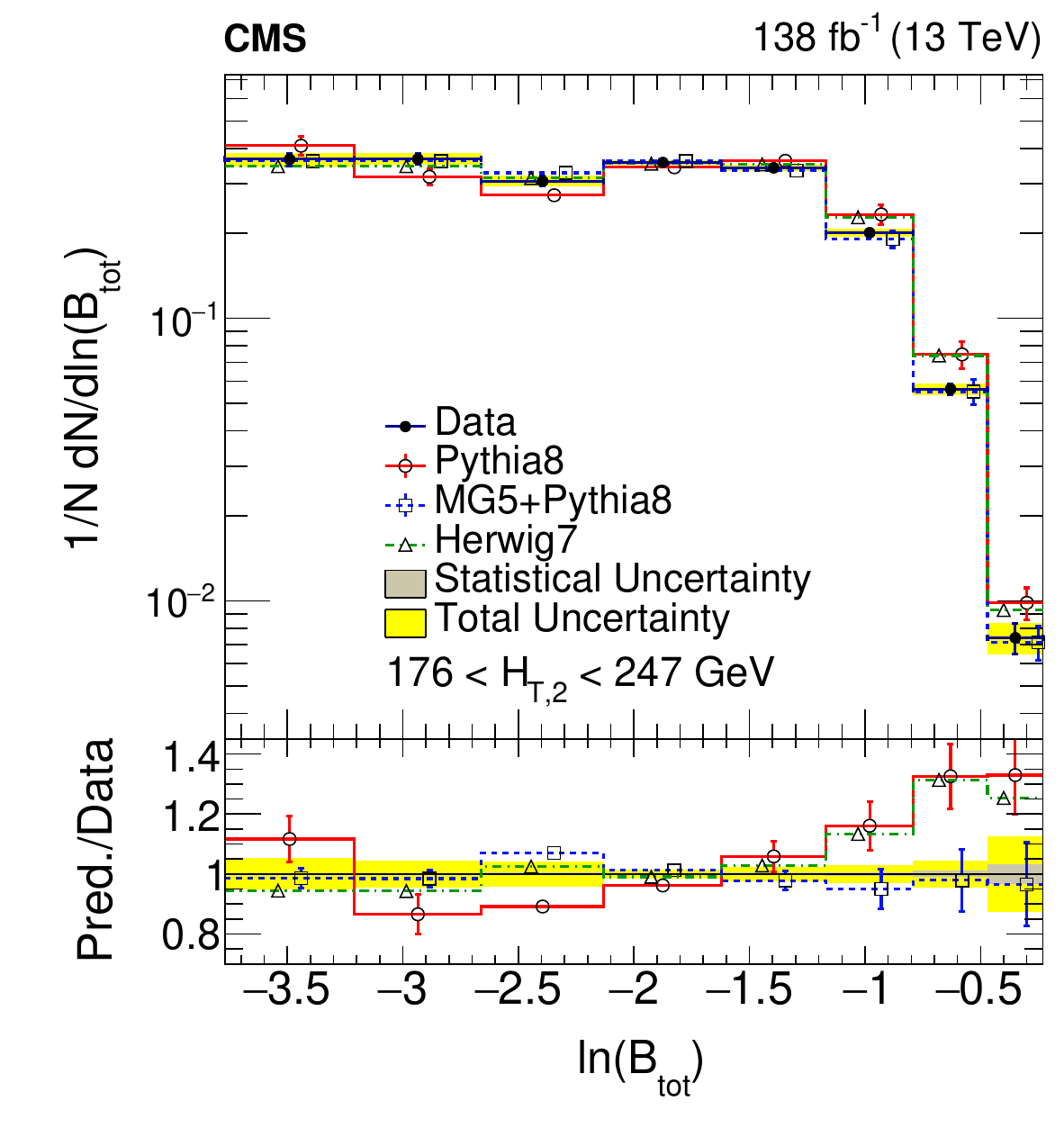}\\
\includegraphics[width=0.32\textwidth]{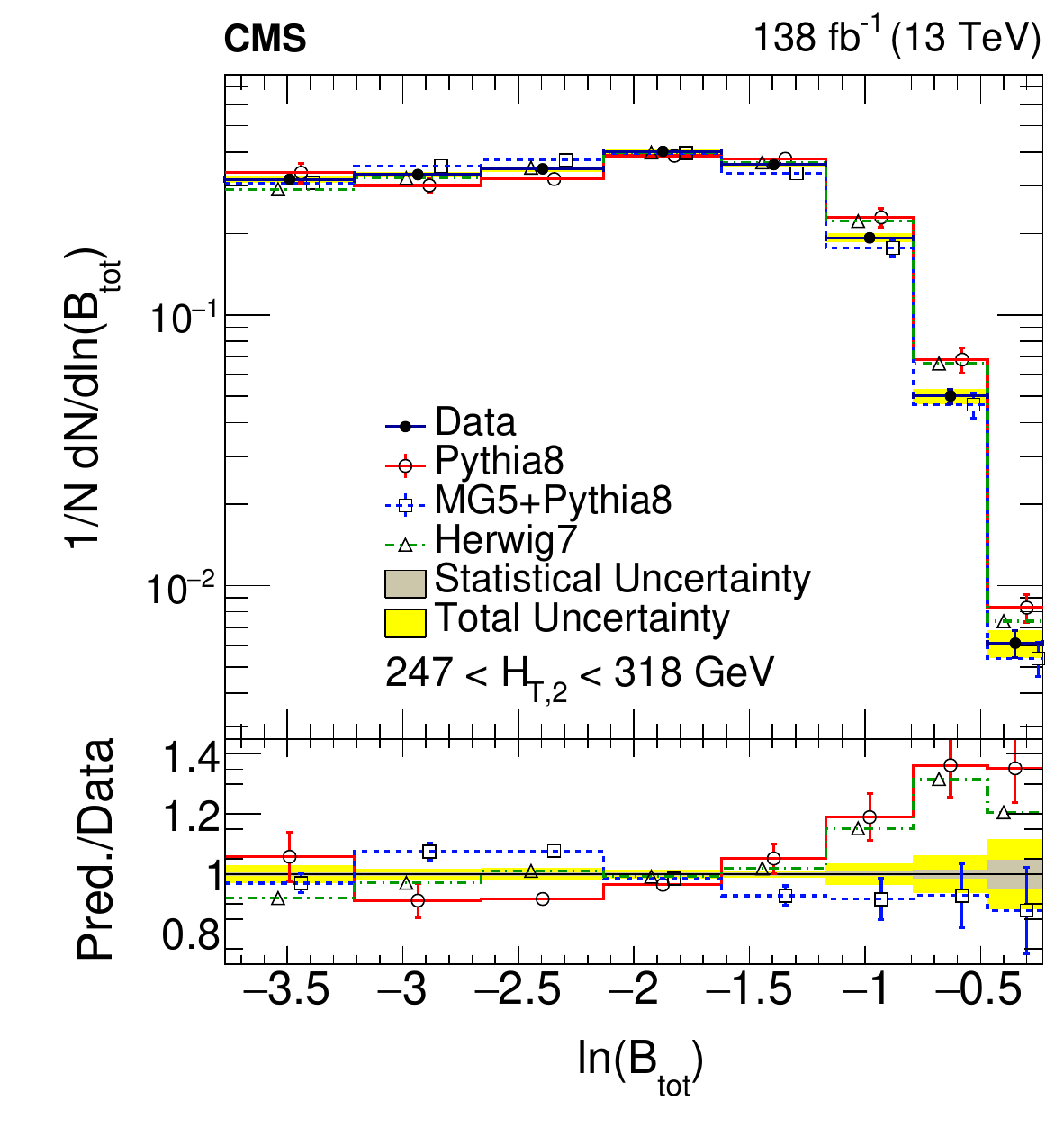}
\includegraphics[width=0.32\textwidth]{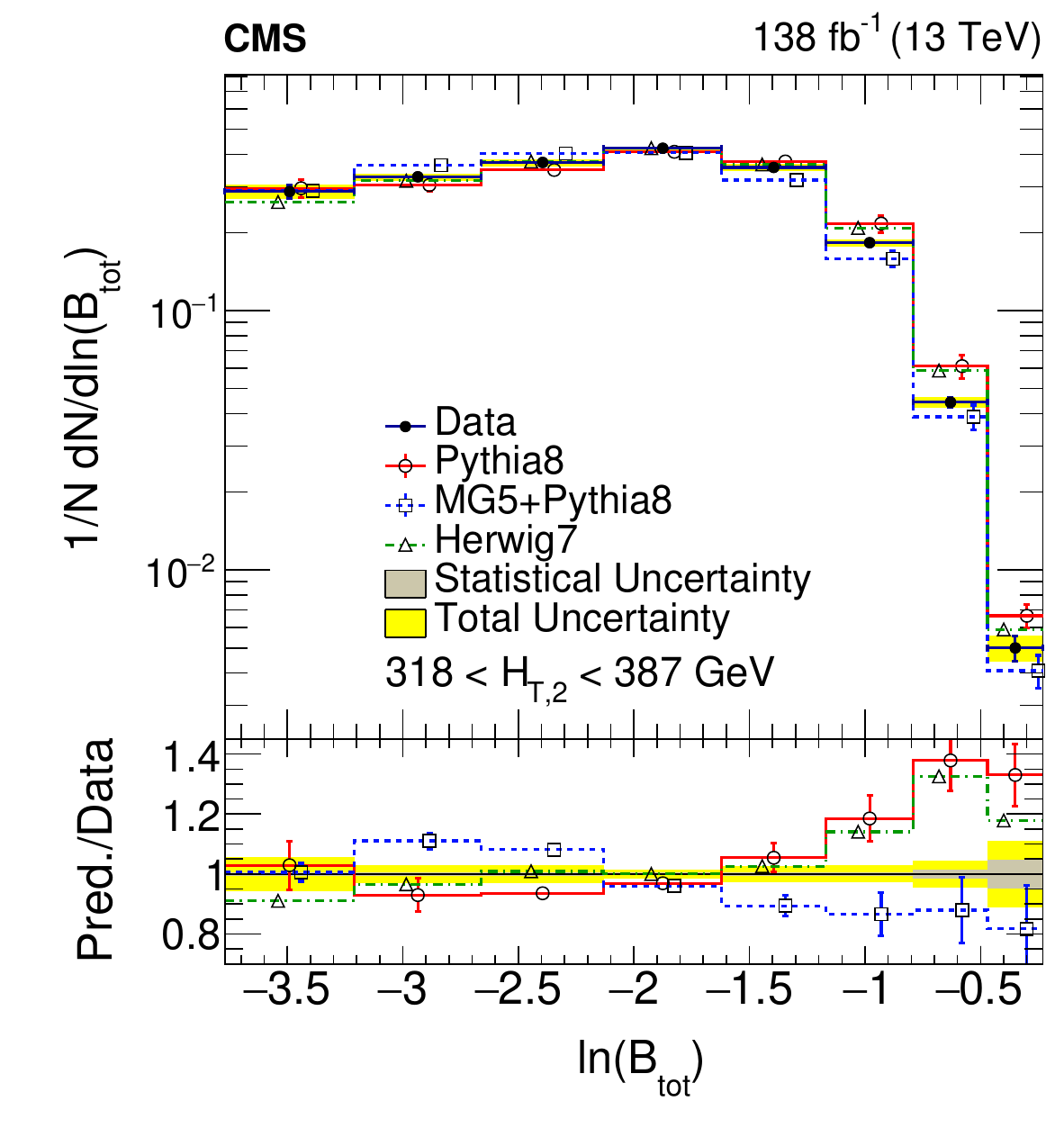}
\includegraphics[width=0.32\textwidth]{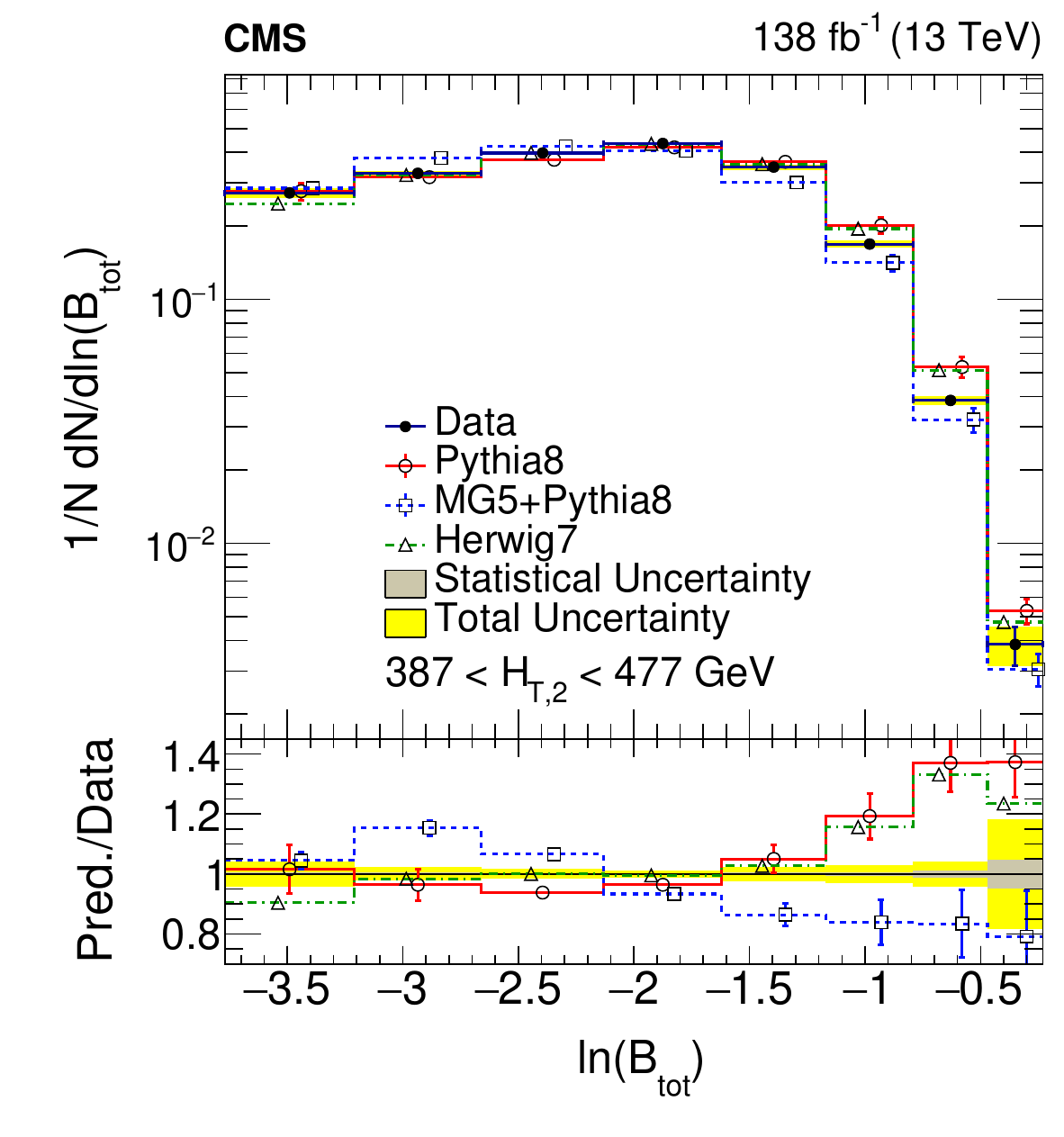}\\
\includegraphics[width=0.32\textwidth]{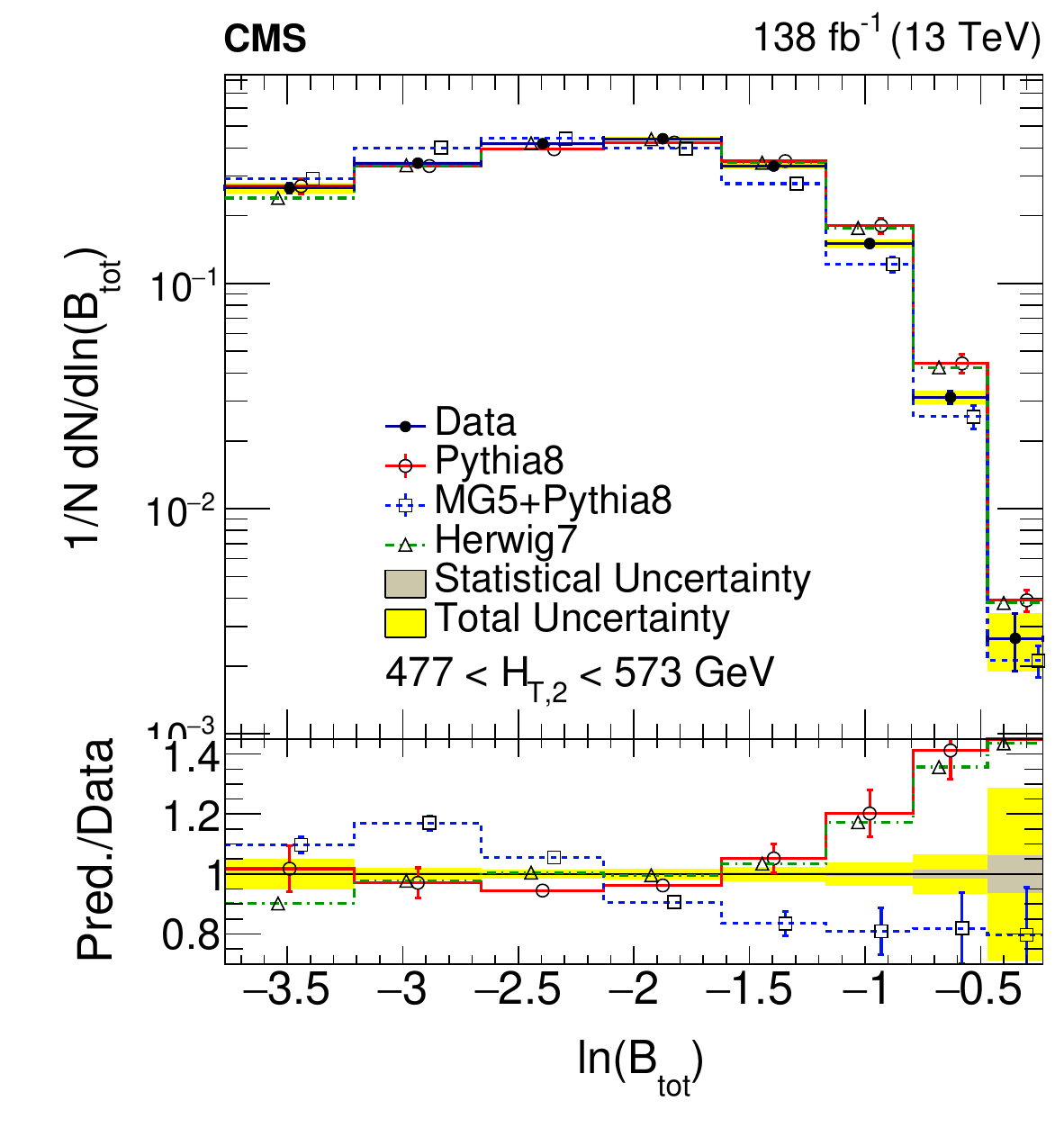}
\includegraphics[width=0.32\textwidth]{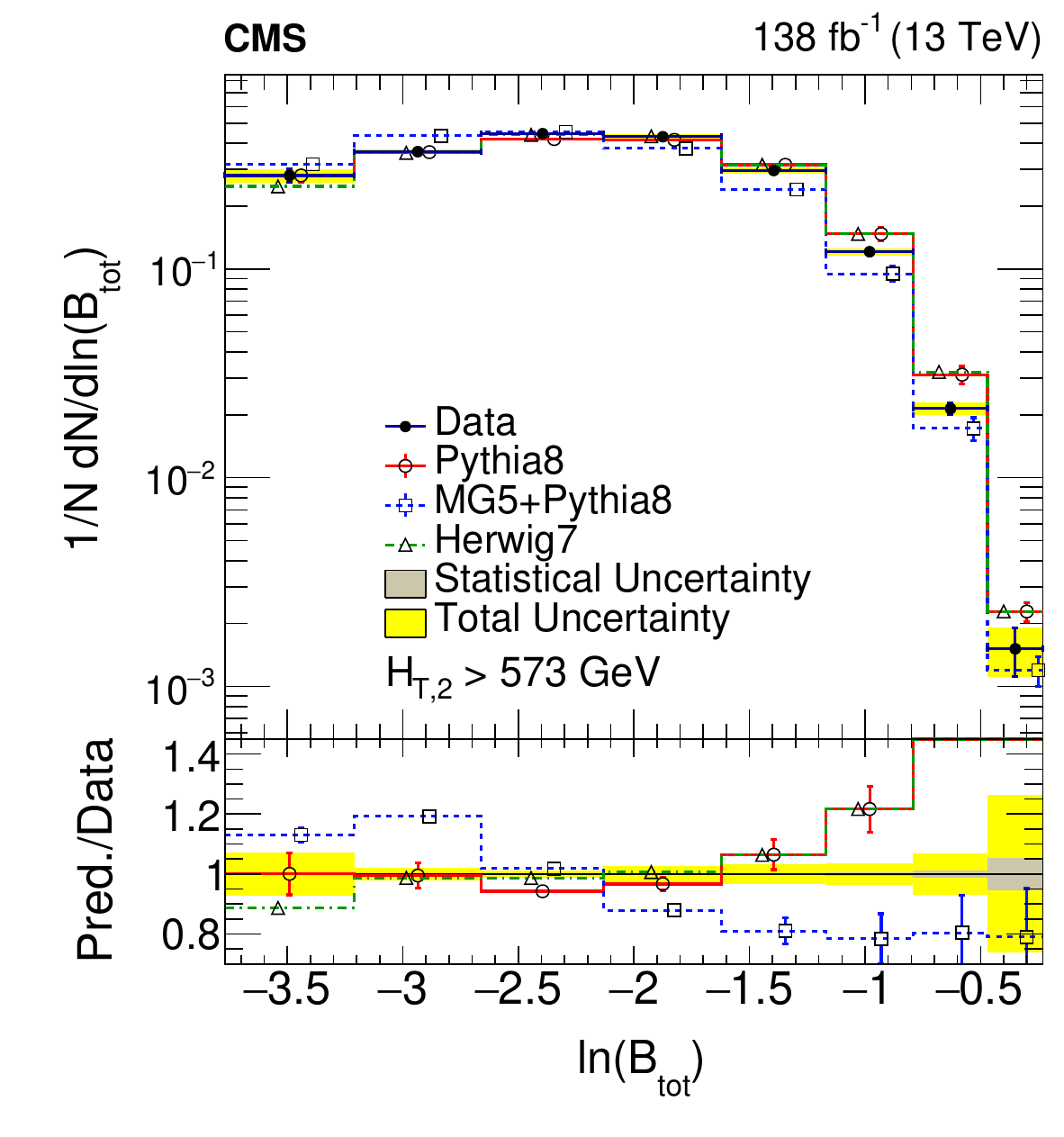}
\caption{Normalized unfolded distributions of total jet broadening ($\bT$) compared with the predictions of \PYTHIAviii (CP5 tune) (red, solid), \HERWIG{}\,7 (CH3 tune) (green, dash-dotted) and \mgpy (blue, dotted) for different $\HTT$ ranges. Notations as in Fig.~\ref{fig:thrustRun2}.}
\label{fig:btRun2}
\end{figure}

\begin{figure}[!htb]
\centering
\includegraphics[width=0.32\textwidth]{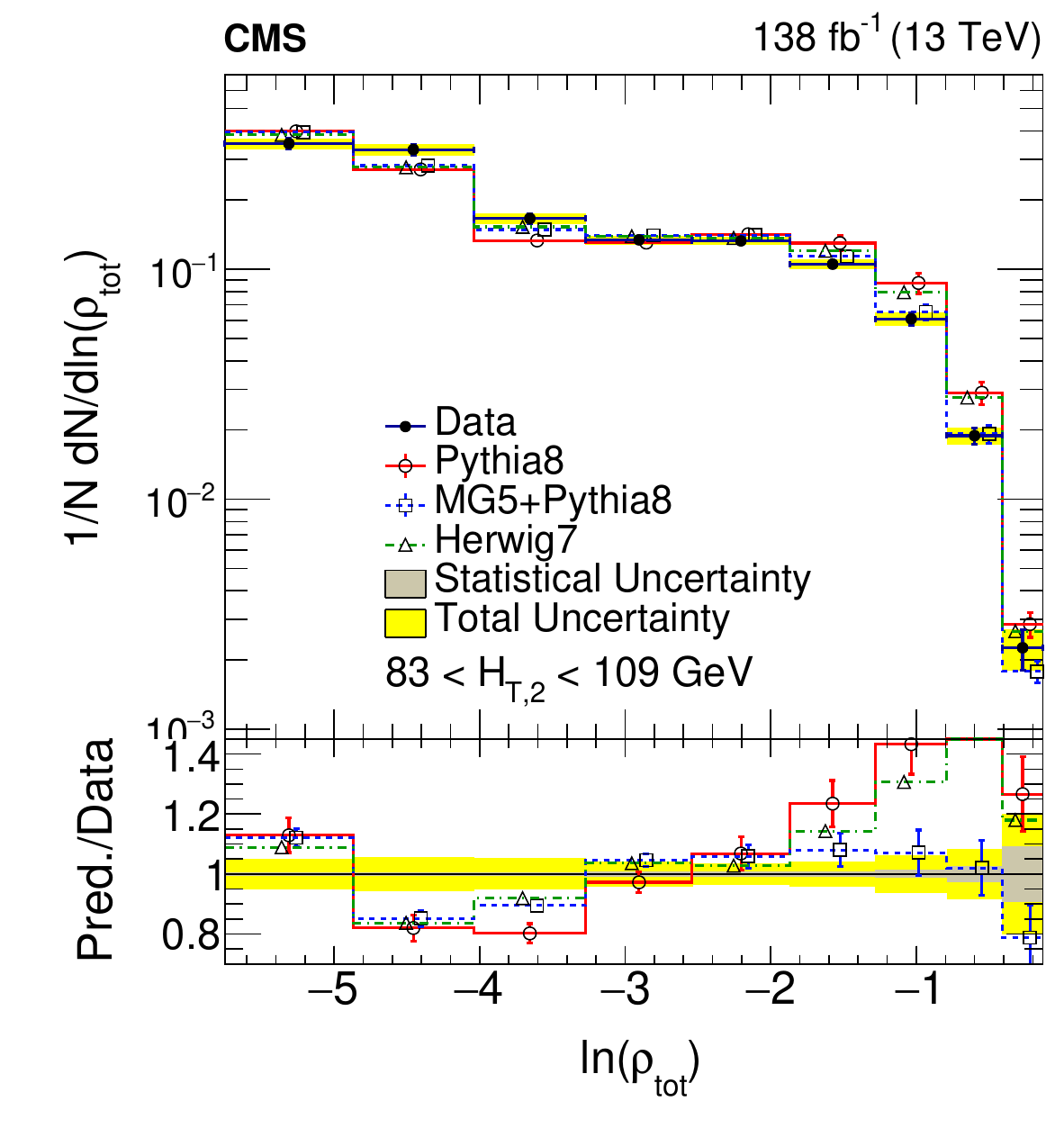}
\includegraphics[width=0.32\textwidth]{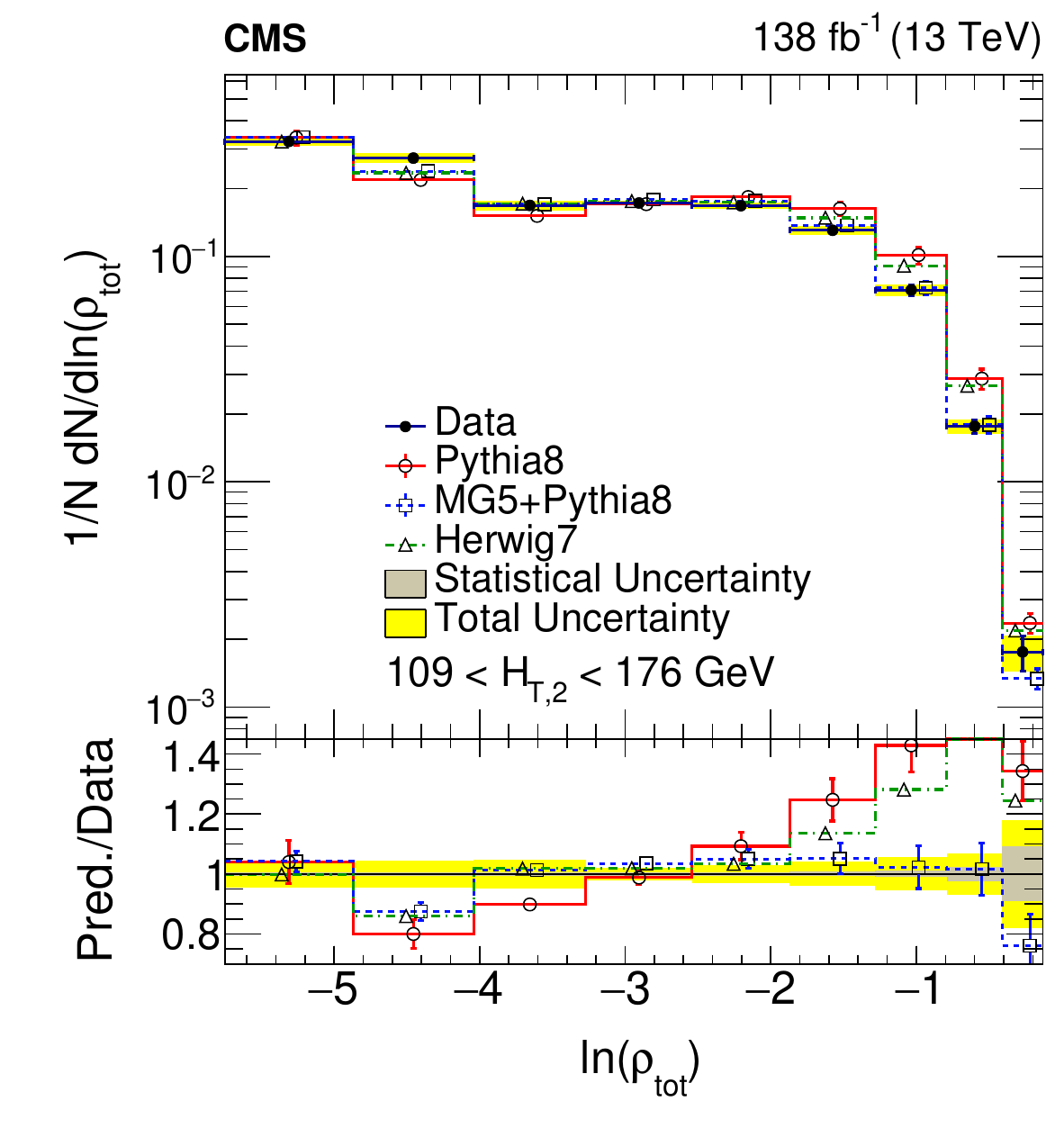}
\includegraphics[width=0.32\textwidth]{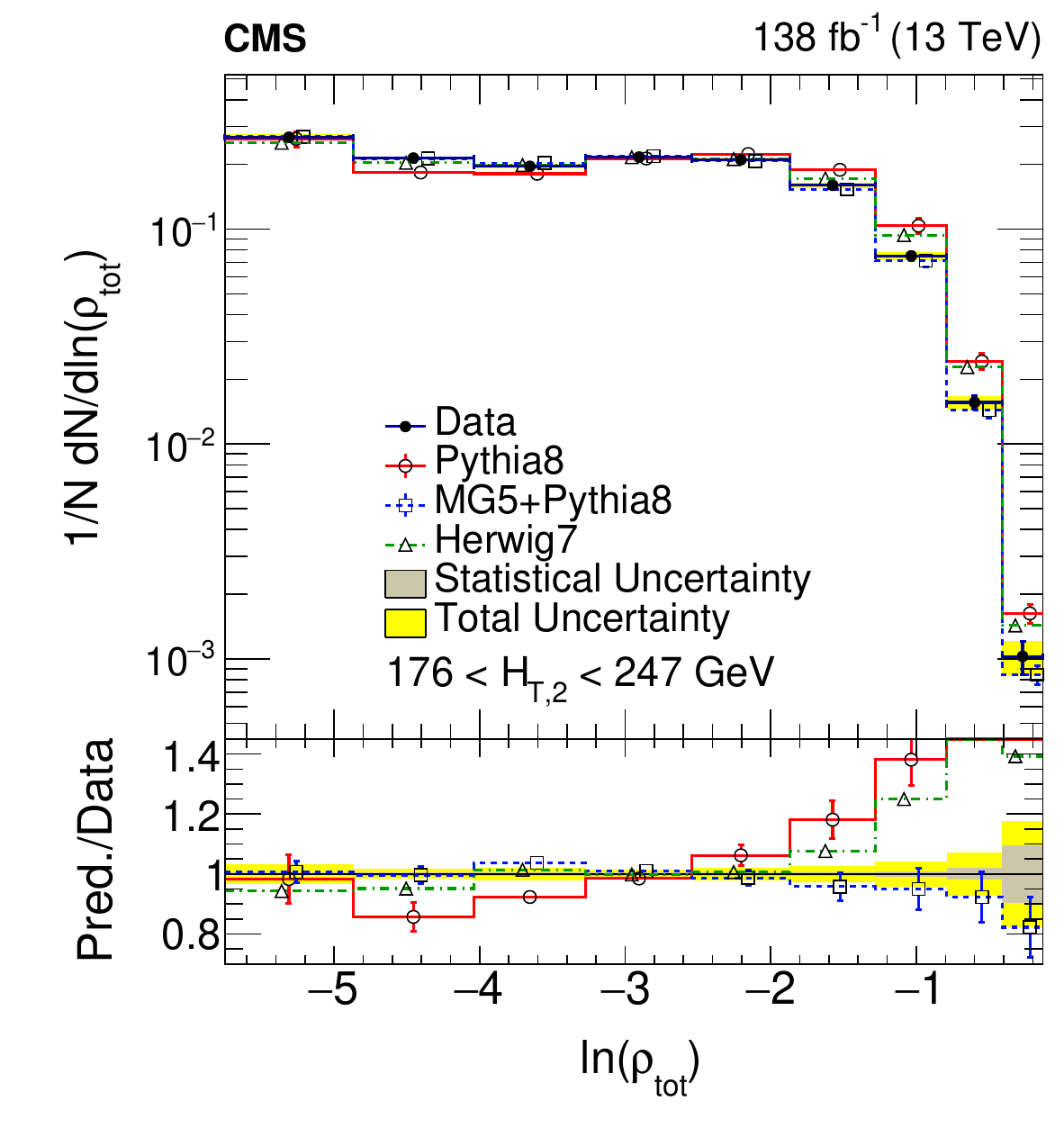}\\
\includegraphics[width=0.32\textwidth]{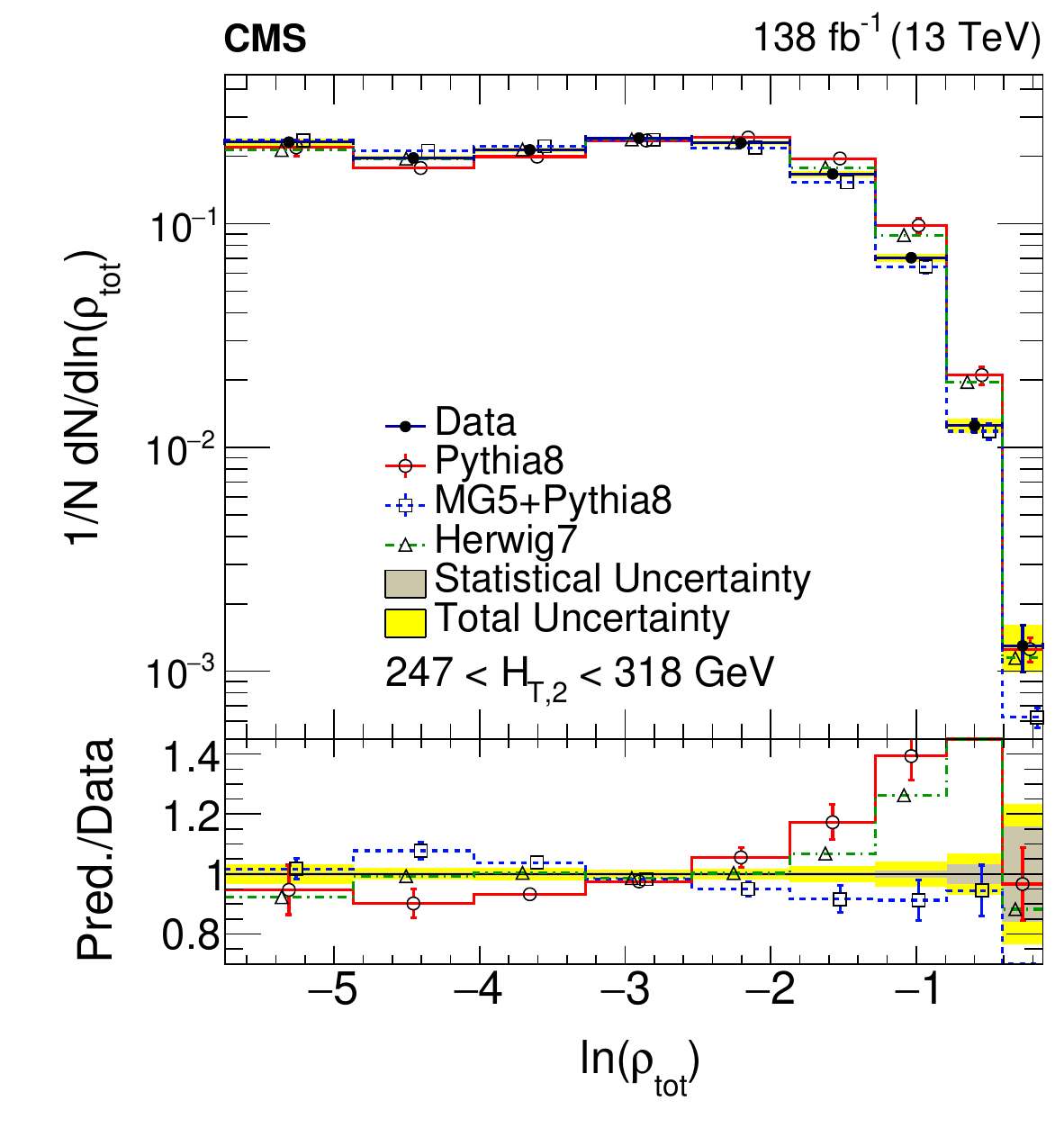}
\includegraphics[width=0.32\textwidth]{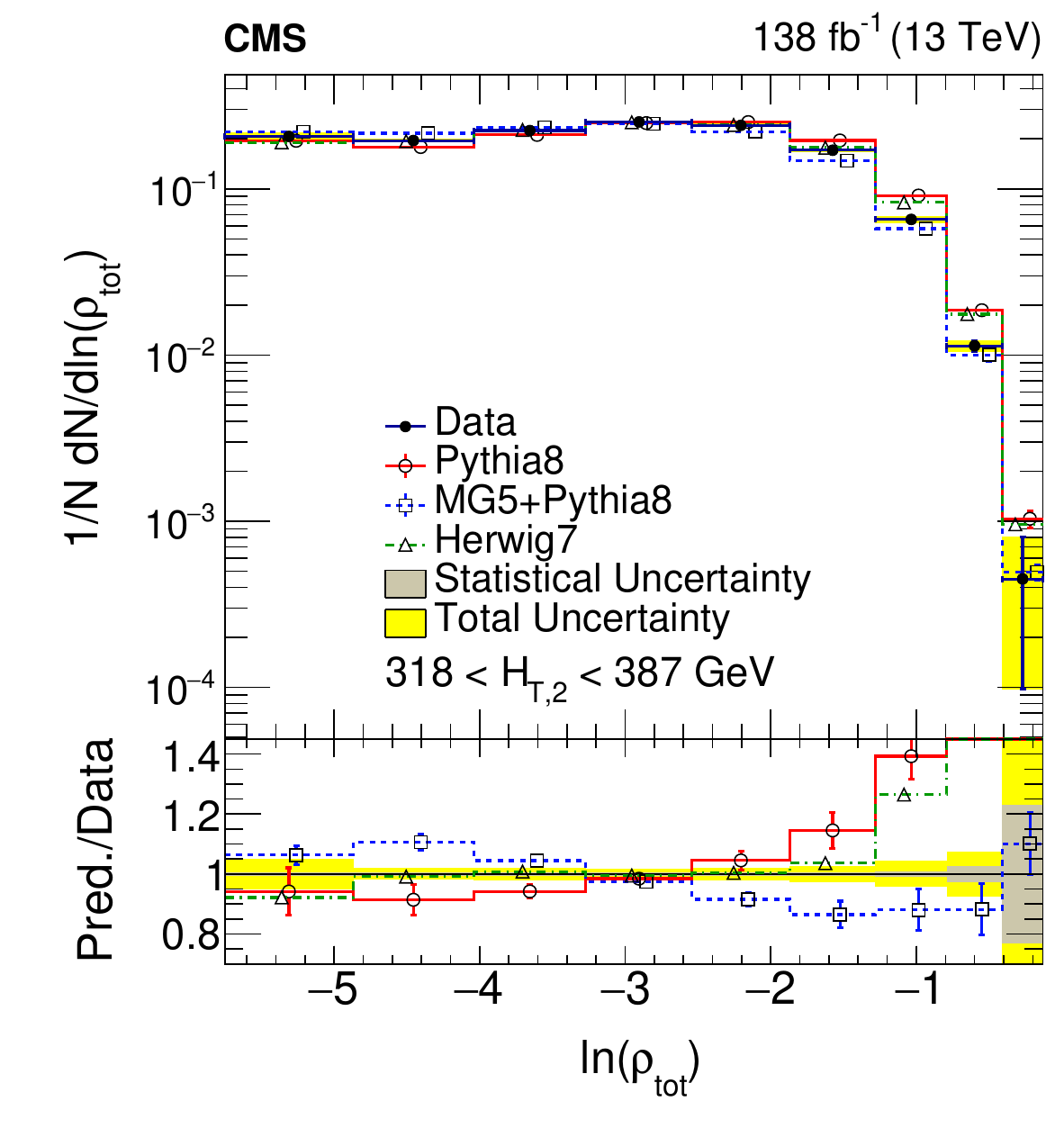}
\includegraphics[width=0.32\textwidth]{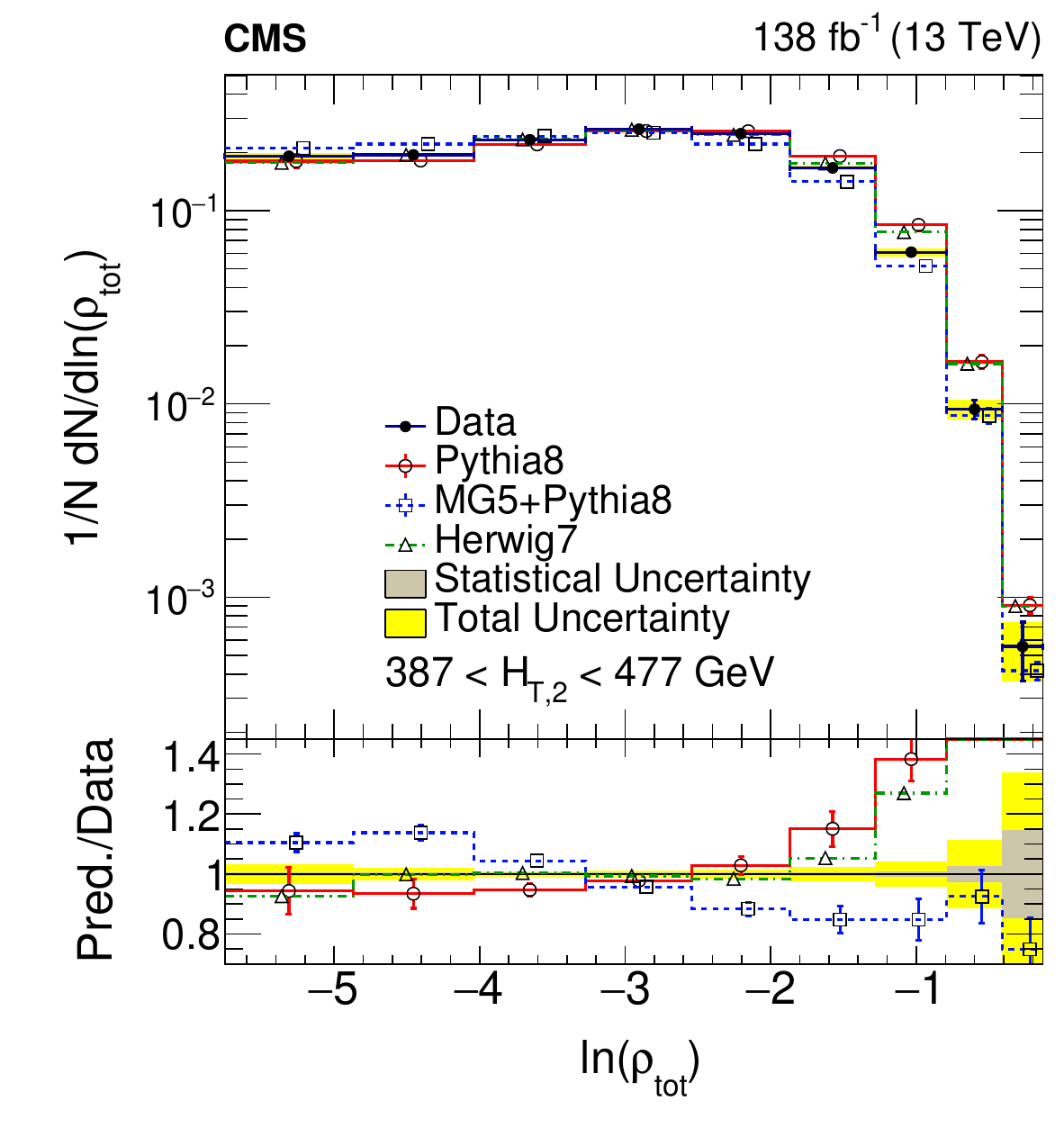}\\
\includegraphics[width=0.32\textwidth]{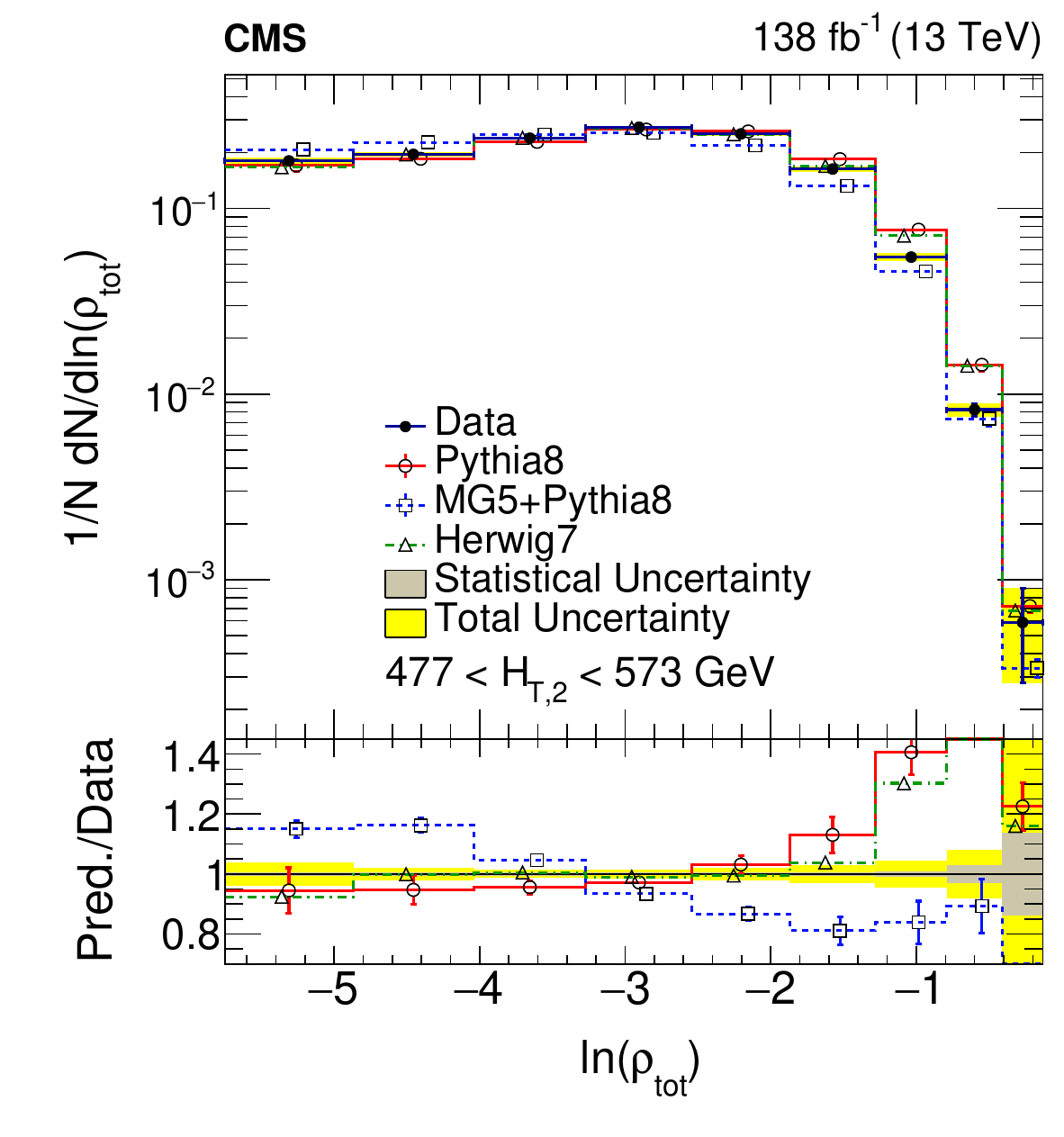}
\includegraphics[width=0.32\textwidth]{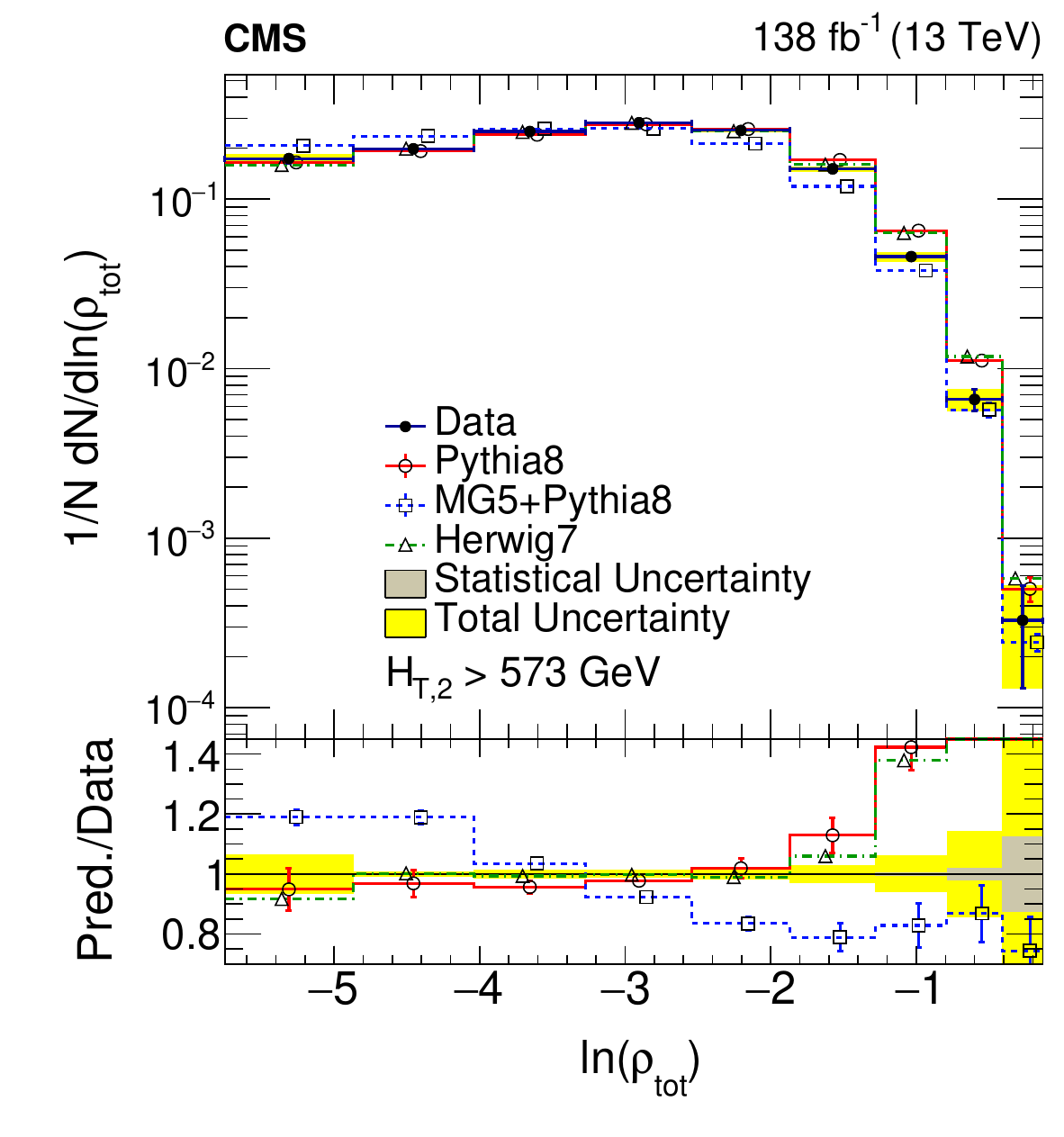}
\caption{Normalized unfolded distributions of total jet mass ($\mass$) compared with the predictions of \PYTHIAviii (CP5 tune) (red, solid), \HERWIG{}\,7 (CH3 tune) (green, dash-dotted) and \mgpy (blue, dotted) for different $\HTT$ ranges. Notations as in Fig.~\ref{fig:thrustRun2}.}
\label{fig:rhoRun2}
\end{figure}

\begin{figure}[!htb]
\centering
\includegraphics[width=0.32\textwidth]{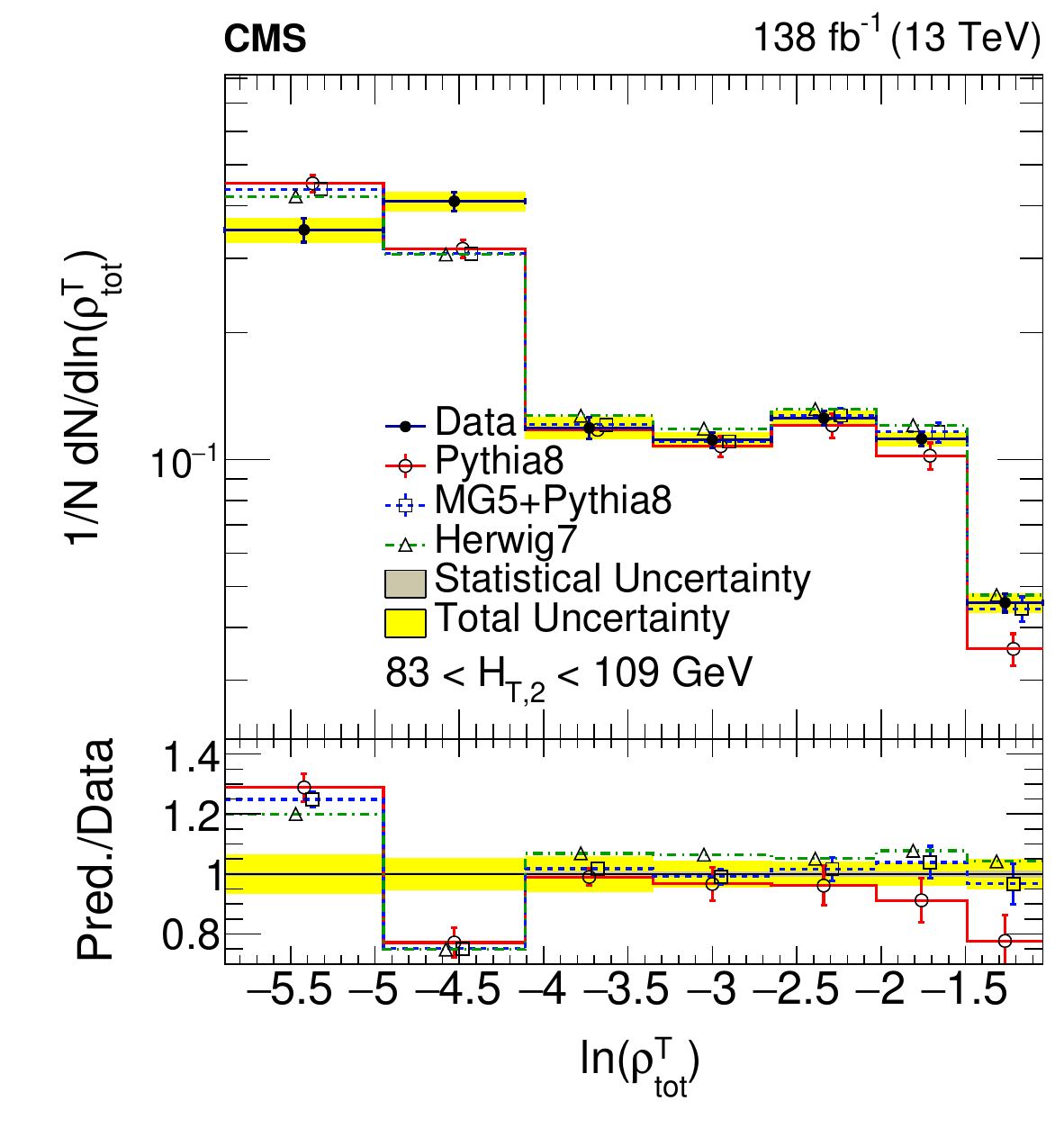}
\includegraphics[width=0.32\textwidth]{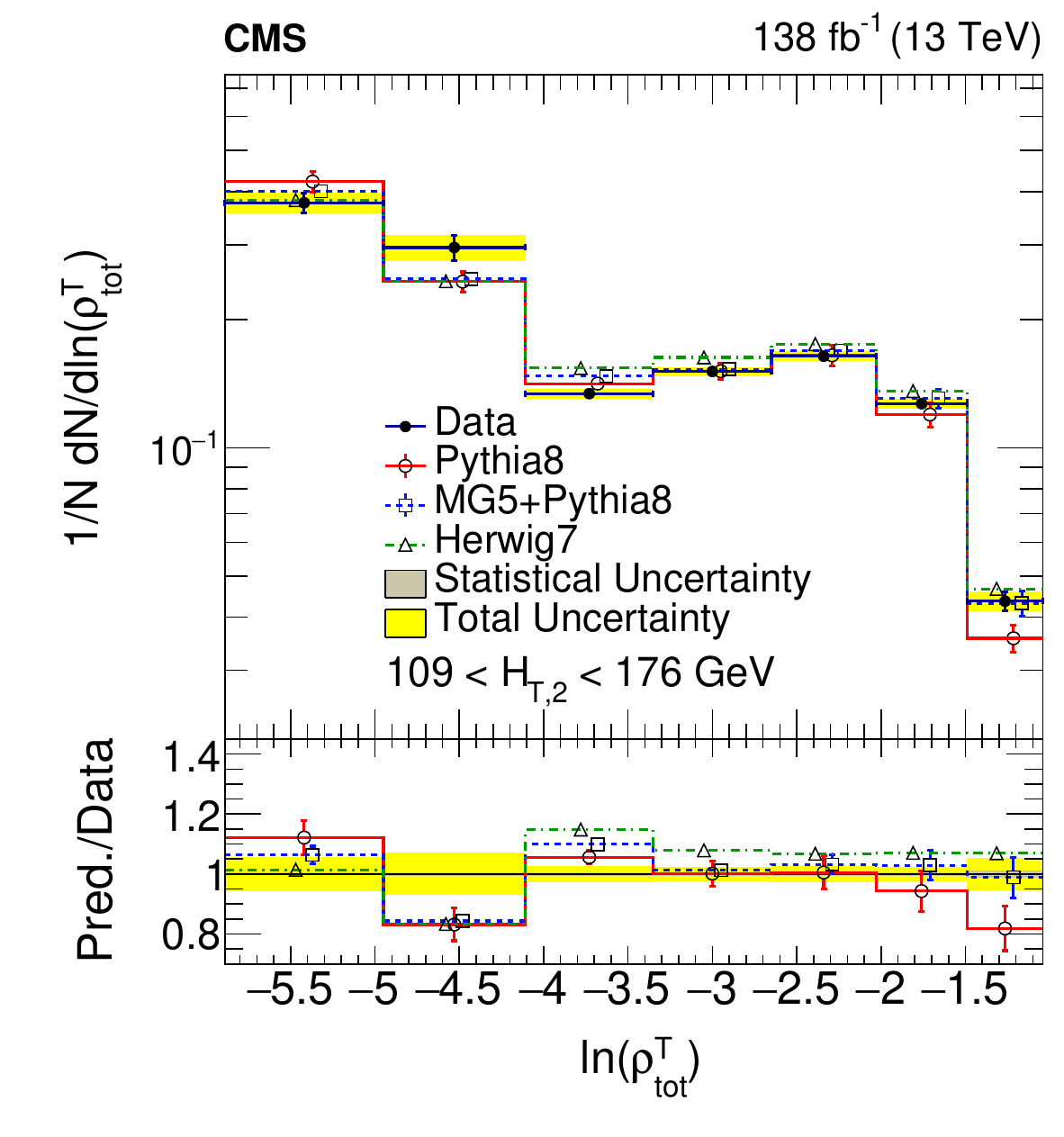}
\includegraphics[width=0.32\textwidth]{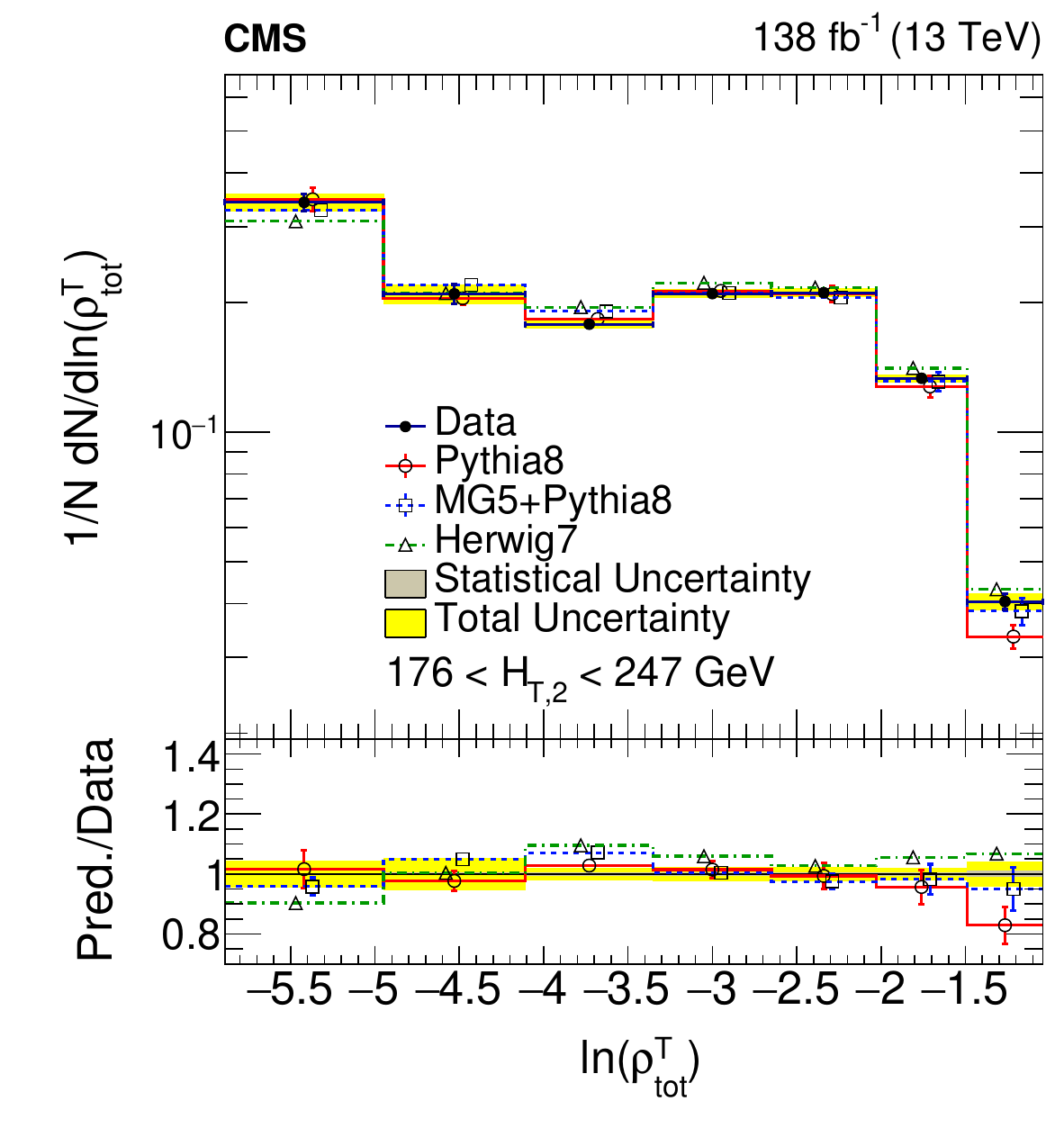}\\
\includegraphics[width=0.32\textwidth]{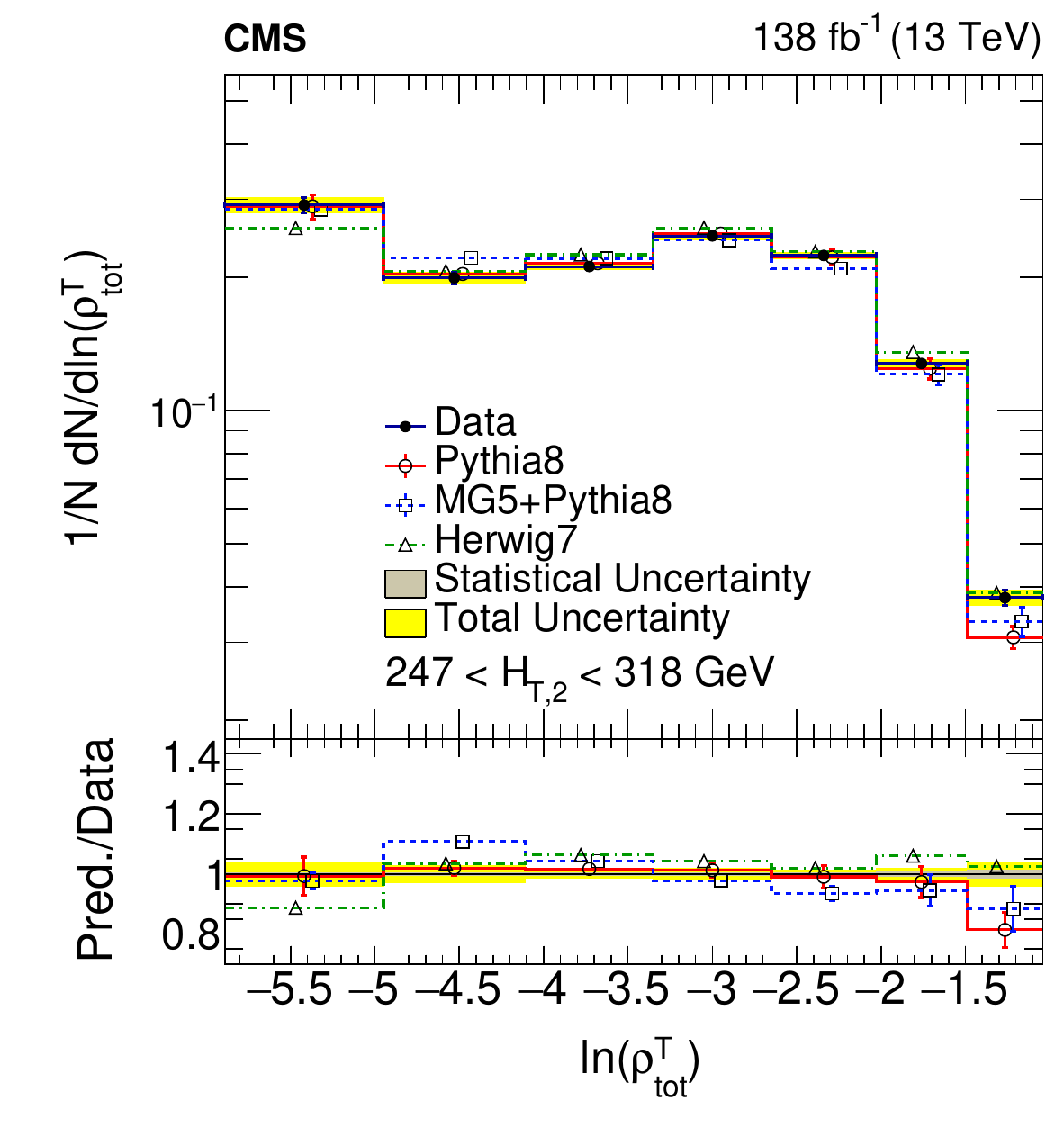}
\includegraphics[width=0.32\textwidth]{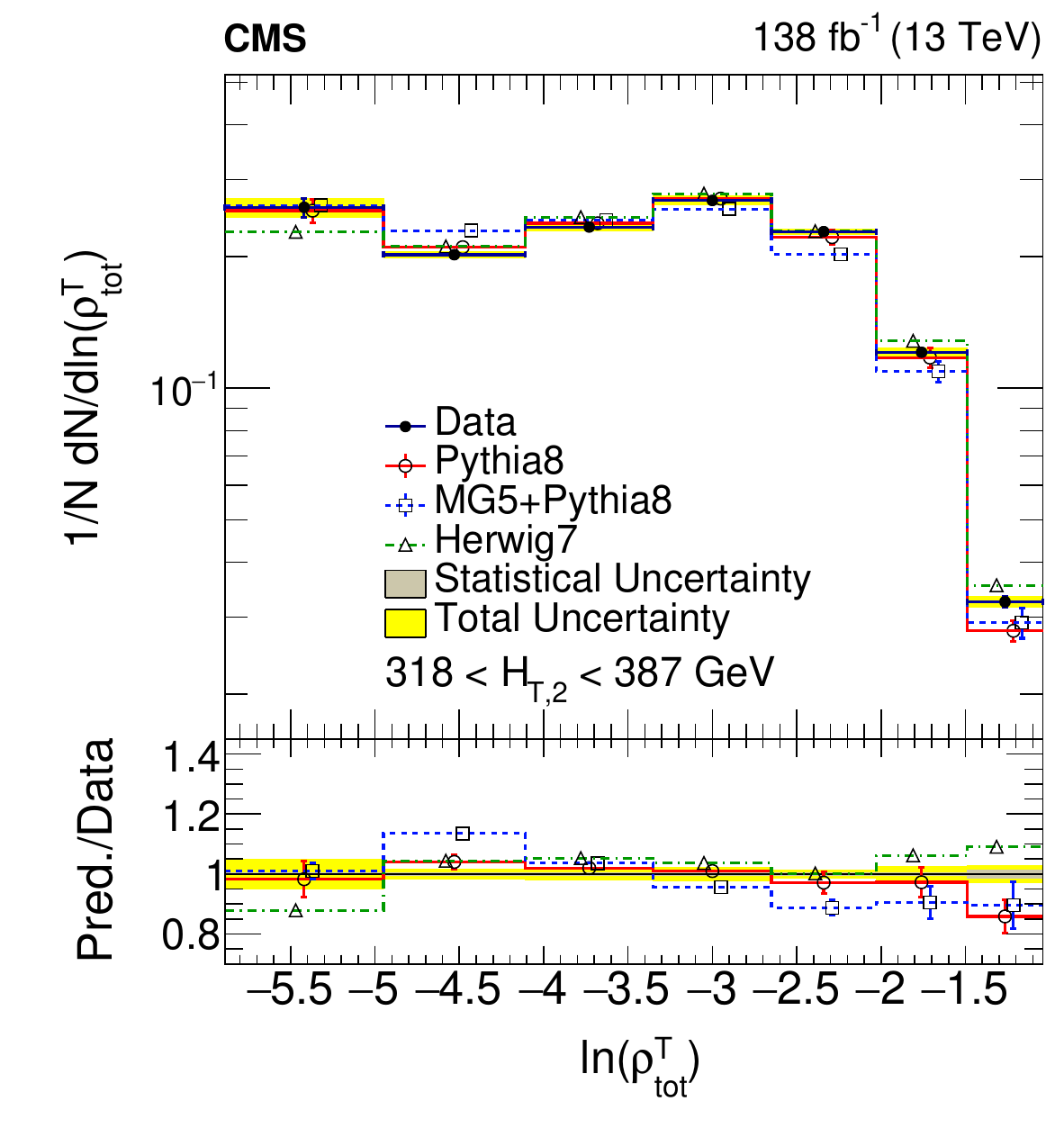}
\includegraphics[width=0.32\textwidth]{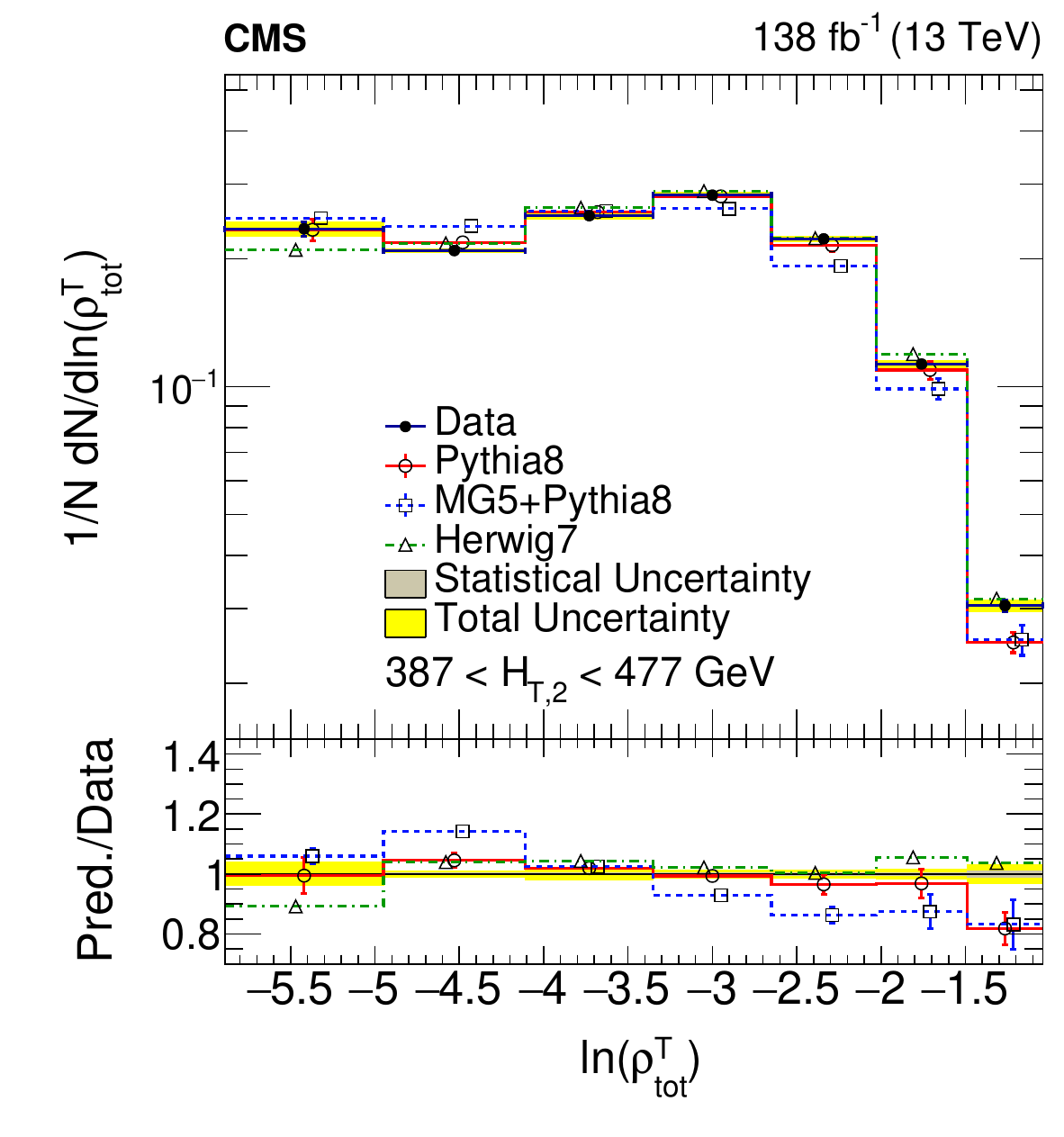}\\
\includegraphics[width=0.32\textwidth]{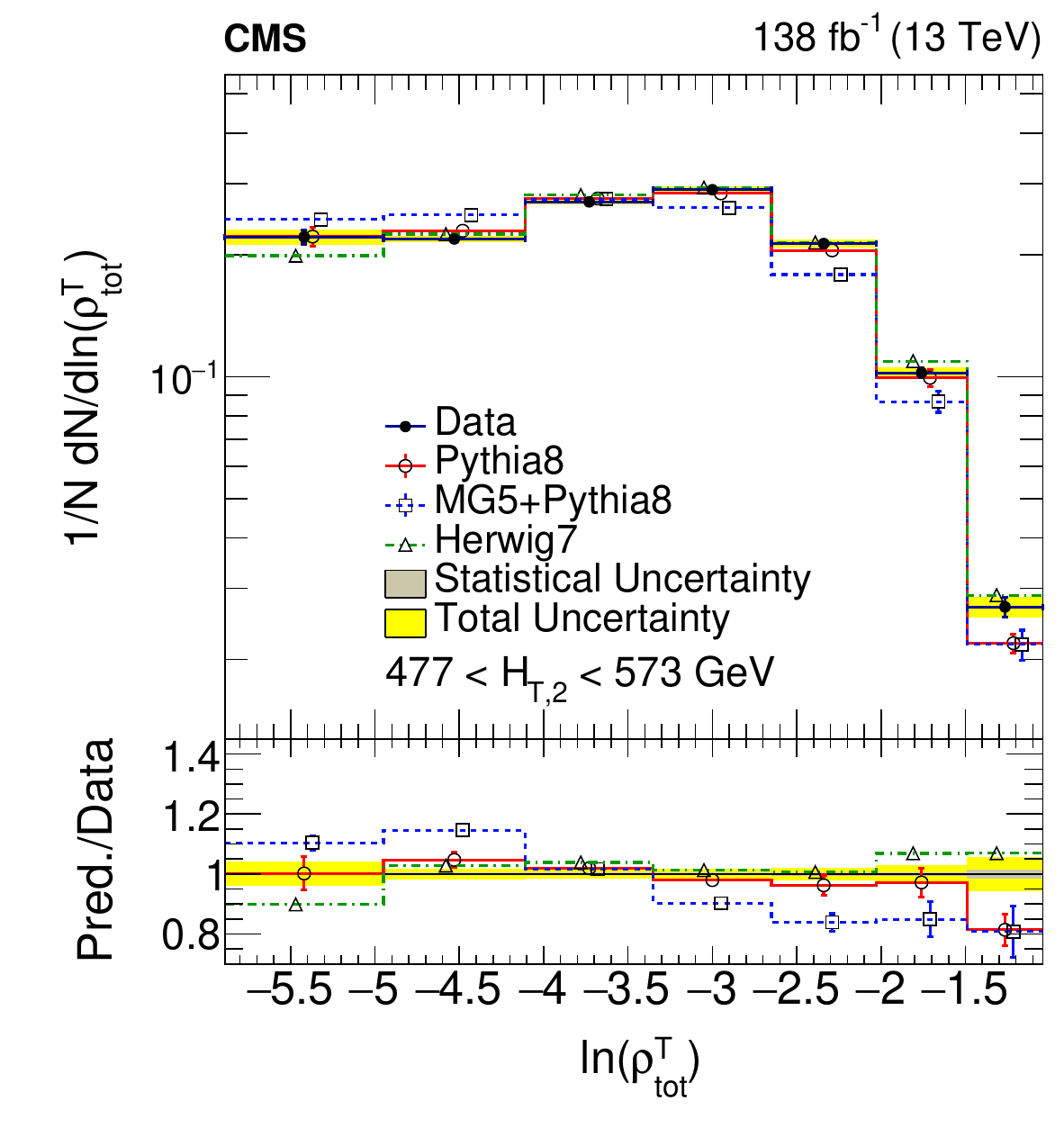}
\includegraphics[width=0.32\textwidth]{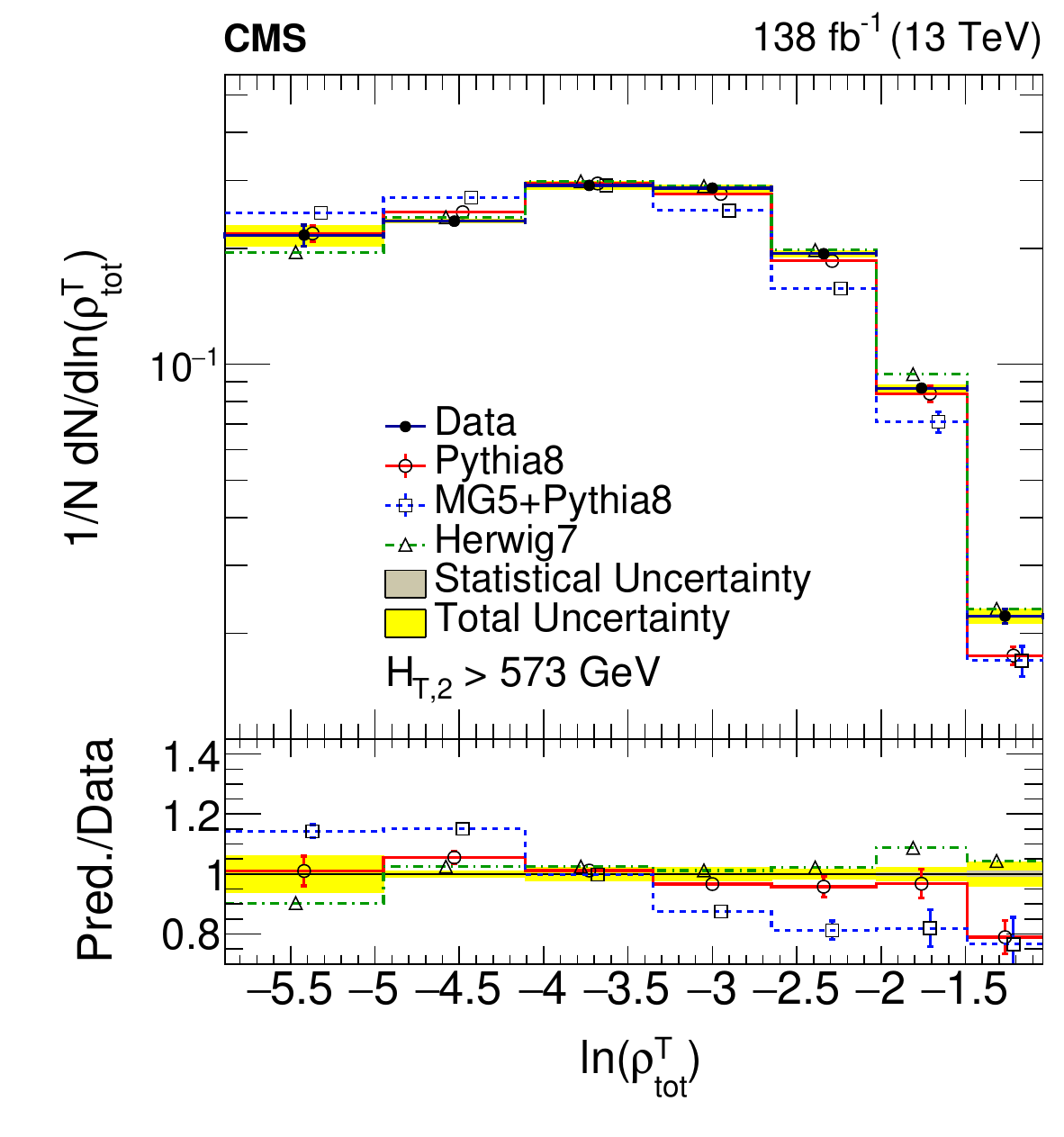}
\caption{Normalized unfolded distributions of total transverse jet mass ($\massp$) compared with the predictions of \PYTHIAviii (CP5 tune) (red, solid), \HERWIG{}\,7 (CH3 tune) (green, dash-dotted) and \mgpy (blue, dotted) for different $\HTT$ ranges. Notations as in Fig.~\ref{fig:thrustRun2}.}
\label{fig:tmassRun2}
\end{figure}

The unfolded distributions of the five ESVs for different $\HTT$ ranges are normalized to unity and shown with the predictions of \PYTHIAviii (CP5 tune), \mgpy (CP5 tune), and \HERWIG{}\,7 (CH3 tune) in Figs.~\ref{fig:thrustRun2}--\ref{fig:tmassRun2}.
In each figure, the inner grey band represents the statistical uncertainty and the yellow band corresponds to the total uncertainty.
The error bars corresponding to the \PYTHIAviii (in red) predictions represent the uncertainties due to variations of the renormalization scale for QCD emissions in ISR and FSR~\cite{Mrenna:2016sih}. 
The error bars corresponding to the \mgpy (in blue) predictions represent the uncertainties due to the renormalization and factorization scale variations~\cite{Butterworth:2015oua}.
The ratios of the predictions of different MC models with observation are shown in the lower panel of each plot. 

Predictions of \PYTHIAviii show good overall agreement with data for $\taup$ and $\massp$ for the events with back-to-back jets topologies as well as multijet, spherical events. 
For $\mass$, $\ytwo$, and $\bT$, \PYTHIAviii overestimates multijet, spherical events. 
The agreement between data and \PYTHIAviii improves with increasing $\HTT$ for all the ESVs. 
The agreement between data and \HERWIG{}\,7 is mostly similar to \PYTHIAviii. 
For lower values of $\HTT$, the agreement between \mgpy predictions and data is good. 
For higher values of $\HTT$, \mgpy overestimates the share of 2-jet events and correspondingly underestimates the fraction of multijet, spherical events for all the ESVs except $\ytwo$. 
The $\ytwo$ depends on the relative value of \pt of the third jet with respect to the two highest \pt jets in the event and is less sensitive to the hadronization of the partons or underlying events.

We note that both \PYTHIAviii and \HERWIG{}\,7, show good agreement with data for $\taup$ and $\massp$, which use \pt of the particles but not for $\mass$, $\bT$, and $\ytwo$ which use the complete momentum information. 
This possibly indicates that there is a shortcoming in the modelling of fragmentation and hadronization. 
On the other hand, \mgpy mostly underestimates multijet, spherical events, and this possibly indicates issues with the combination of ME calculations with the PS. 

\section{Summary}\label{sec:summary}
In this study the charged particles inside jets have been used to evaluate five event shape variables.
The complement of transverse thrust , total jet mass, third jet resolution parameter, total jet broadening, and total transverse jet mass are studied for different ranges of $\HTT$, the average \pt of the leading and the second leading jets in an event. Data from proton-proton collisions at $\sqrt{s} = 13\TeV$, collected during 2016--2018, have been used for the study. For each event shape variable, detector effects are corrected using a two-dimensional unfolding procedure in which each ESV is unfolded simultaneously with $\HTT$. Detailed comparisons of the leading-order predictions of \PYTHIAviii (CP5 tune), {\MGvATNLO}+\PYTHIAviii (CP5 tune), and \HERWIG{}\,7 (CH3 tune) with data have been made, covering a large phase space. With the increase in the energy of the parton-parton scattering, the agreement improves for \PYTHIAviii, remains unchanged for \HERWIG{}\,7, and worsens for {\MGvATNLO}+\PYTHIAviii. Overall, the results indicate that our understanding of the energy flow in an event at different energy scales needs improvement.


\clearpage
\begin{acknowledgments}
We congratulate our colleagues in the CERN accelerator departments for the excellent performance of the LHC and thank the technical and administrative staffs at CERN and at other CMS institutes for their contributions to the success of the CMS effort. In addition, we gratefully acknowledge the computing centres and personnel of the Worldwide LHC Computing Grid and other centres for delivering so effectively the computing infrastructure essential to our analyses. Finally, we acknowledge the enduring support for the construction and operation of the LHC, the CMS detector, and the supporting computing infrastructure provided by the following funding agencies: SC (Armenia), BMBWF and FWF (Austria); FNRS and FWO (Belgium); CNPq, CAPES, FAPERJ, FAPERGS, and FAPESP (Brazil); MES and BNSF (Bulgaria); CERN; CAS, MoST, and NSFC (China); MINCIENCIAS (Colombia); MSES and CSF (Croatia); RIF (Cyprus); SENESCYT (Ecuador); ERC PRG and PSG, TARISTU24-TK10 and MoER TK202 (Estonia); Academy of Finland, MEC, and HIP (Finland); CEA and CNRS/IN2P3 (France); SRNSF (Georgia); BMFTR, DFG, and HGF (Germany); GSRI (Greece); NKFIH (Hungary); DAE and DST (India); IPM (Iran); SFI (Ireland); INFN (Italy); MSIT and NRF (Republic of Korea); MES (Latvia); LMTLT (Lithuania); MOE and UM (Malaysia); BUAP, CINVESTAV, CONACYT, LNS, SEP, and UASLP-FAI (Mexico); MOS (Montenegro); MBIE (New Zealand); PAEC (Pakistan); MES, NSC, and NAWA (Poland); FCT (Portugal);  MESTD (Serbia); MICIU/AEI and PCTI (Spain); MOSTR (Sri Lanka); Swiss Funding Agencies (Switzerland); MST (Taipei); MHESI (Thailand); TUBITAK and TENMAK (T\"{u}rkiye); NASU (Ukraine); STFC (United Kingdom); DOE and NSF (USA).

\hyphenation{Rachada-pisek} Individuals have received support from the Marie-Curie programme and the European Research Council and Horizon 2020 Grant, contract Nos.\ 675440, 724704, 752730, 758316, 765710, 824093, 101115353, 101002207, 101001205, and COST Action CA16108 (European Union); the Leventis Foundation; the Alfred P.\ Sloan Foundation; the Alexander von Humboldt Foundation; the Science Committee, project no. 22rl-037 (Armenia); the Fonds pour la Formation \`a la Recherche dans l'Industrie et dans l'Agriculture (FRIA) and Fonds voor Wetenschappelijk Onderzoek contract No. 1228724N (Belgium); the Beijing Municipal Science \& Technology Commission, No. Z191100007219010, the Fundamental Research Funds for the Central Universities, the Ministry of Science and Technology of China under Grant No. 2023YFA1605804, the Natural Science Foundation of China under Grant No. 12535004, and USTC Research Funds of the Double First-Class Initiative No.\ YD2030002017 (China); the Ministry of Education, Youth and Sports (MEYS) of the Czech Republic; the Shota Rustaveli National Science Foundation, grant FR-22-985 (Georgia); the Deutsche Forschungsgemeinschaft (DFG), among others, under Germany's Excellence Strategy -- EXC 2121 ``Quantum Universe" -- 390833306, and under project number 400140256 - GRK2497; the Hellenic Foundation for Research and Innovation (HFRI), Project Number 2288 (Greece); the Hungarian Academy of Sciences, the New National Excellence Program - \'UNKP, the NKFIH research grants K 131991, K 133046, K 138136, K 143460, K 143477, K 146913, K 146914, K 147048, 2020-2.2.1-ED-2021-00181, TKP2021-NKTA-64, and 2025-1.1.5-NEMZ\_KI-2025-00004 (Hungary); the Council of Science and Industrial Research, India; ICSC -- National Research Centre for High Performance Computing, Big Data and Quantum Computing, FAIR -- Future Artificial Intelligence Research, and CUP I53D23001070006 (Mission 4 Component 1), funded by the NextGenerationEU program (Italy); the Latvian Council of Science; the Ministry of Education and Science, project no. 2022/WK/14, and the National Science Center, contracts Opus 2021/41/B/ST2/01369, 2021/43/B/ST2/01552, 2023/49/B/ST2/03273, and the NAWA contract BPN/PPO/2021/1/00011 (Poland); the Funda\c{c}\~ao para a Ci\^encia e a Tecnologia (Portugal); the National Priorities Research Program by Qatar National Research Fund; MICIU/AEI/10.13039/501100011033, ERDF/EU, ``European Union NextGenerationEU/PRTR", and Programa Severo Ochoa del Principado de Asturias (Spain); the Chulalongkorn Academic into Its 2nd Century Project Advancement Project, the National Science, Research and Innovation Fund program IND\_FF\_68\_369\_2300\_097, and the Program Management Unit for Human Resources \& Institutional Development, Research and Innovation, grant B39G680009 (Thailand); the Eric \& Wendy Schmidt Fund for Strategic Innovation through the CERN Next Generation Triggers project under grant agreement number SIF-2023-004; the Kavli Foundation; the Nvidia Corporation; the SuperMicro Corporation; the Welch Foundation, contract C-1845; and the Weston Havens Foundation (USA).
\end{acknowledgments}\section*{Data availability} Release and preservation of data used by the CMS Collaboration as the basis for publications is guided by the  \href{https://doi.org/10.7483/OPENDATA.CMS.1BNU.8V1W}{CMS data preservation, re-use and open access policy}.

\bibliography{auto_generated}
\cleardoublepage \appendix\section{The CMS Collaboration \label{app:collab}}\begin{sloppypar}\hyphenpenalty=5000\widowpenalty=500\clubpenalty=5000\cmsinstitute{Yerevan Physics Institute, Yerevan, Armenia}
{\tolerance=6000
A.~Hayrapetyan, V.~Makarenko\cmsorcid{0000-0002-8406-8605}, A.~Tumasyan\cmsAuthorMark{1}\cmsorcid{0009-0000-0684-6742}
\par}
\cmsinstitute{Institut f\"{u}r Hochenergiephysik, Vienna, Austria}
{\tolerance=6000
W.~Adam\cmsorcid{0000-0001-9099-4341}, J.W.~Andrejkovic, L.~Benato\cmsorcid{0000-0001-5135-7489}, T.~Bergauer\cmsorcid{0000-0002-5786-0293}, K.~Damanakis\cmsorcid{0000-0001-5389-2872}, M.~Dragicevic\cmsorcid{0000-0003-1967-6783}, C.~Giordano\cmsorcid{0000-0001-6317-2481}, P.S.~Hussain\cmsorcid{0000-0002-4825-5278}, M.~Jeitler\cmsAuthorMark{2}\cmsorcid{0000-0002-5141-9560}, N.~Krammer\cmsorcid{0000-0002-0548-0985}, A.~Li\cmsorcid{0000-0002-4547-116X}, D.~Liko\cmsorcid{0000-0002-3380-473X}, I.~Mikulec\cmsorcid{0000-0003-0385-2746}, J.~Schieck\cmsAuthorMark{2}\cmsorcid{0000-0002-1058-8093}, R.~Sch\"{o}fbeck\cmsAuthorMark{2}\cmsorcid{0000-0002-2332-8784}, D.~Schwarz\cmsorcid{0000-0002-3821-7331}, M.~Shooshtari\cmsorcid{0009-0004-8882-4887}, M.~Sonawane\cmsorcid{0000-0003-0510-7010}, W.~Waltenberger\cmsorcid{0000-0002-6215-7228}, C.-E.~Wulz\cmsAuthorMark{2}\cmsorcid{0000-0001-9226-5812}
% ...
\par}
\vskip\cmsinstskip
\dag:~Deceased\\
$^{1}$Also at Yerevan State University, Yerevan, Armenia\\
$^{2}$Also at TU Wien, Vienna, Austria\\
% ...
\end{sloppypar}
%%% END EDITABLE REGION %%%
% skeleton_end
\end{document}